\documentclass[]{imsart}
\usepackage[sectionbib]{natbib}
\usepackage{array,epsfig,fancyheadings,rotating}
\usepackage[]{hyperref}  

\usepackage{siunitx}
\usepackage{amsthm}
\usepackage{amsmath}
\usepackage{amssymb, amsfonts}
\usepackage{longtable}
\usepackage{bm}
\usepackage{graphicx}
\usepackage{color}
\usepackage{floatrow}
\usepackage{units}
\usepackage{url}
\usepackage{verbatim}
\usepackage{epstopdf}
\usepackage{booktabs,caption,fixltx2e}

\usepackage[expand]{gettitlestring}
\makeatletter
\GetTitleStringDisableCommands{%
  \let\enit@align\@firstofone
  \let\enit@format\@firstofone
}
\makeatother

\usepackage{placeins} 
\usepackage{nameref}

\doi{10.1214/154957804100000000}
\pubyear{0000}
\volume{0}
\firstpage{1}
\lastpage{0}

\newcommand\plotpath[1] {./#1}

\newcommand{\iid}{\stackrel{\mathrm{iid}}{\sim}}

\newcommand{\bth}{\boldsymbol\theta}
\newcommand{\bka}{\hbox{$\boldsymbol{\kappa$}}}
\newcommand{\bpi}{\hbox{$\boldsymbol{\pi$}}}
\newcommand{\bbe}{\hbox{$\boldsymbol{\beta$}}}
\newcommand{\bx} {\mathbf{x}}
\newcommand{\by} {\mathbf{y}}
\newcommand{\bxy} {\mathbf{u}}
\newcommand{\bxp} {\mathbf{v}}
\newcommand{\ie} {\hbox{\textit{i.e.}}}
\newcommand{\eg} {\hbox{\textit{e.g.}}}
\newcommand{\hy}{g}   
\newcommand{\hmy}{t}    

\newcommand{\is}[1] {#1^{(s)}}
\newcommand{\sampled}[1] {\hbox{$\mathbf{ \boldsymbol{#1}}^{(s)}$}}
\newcommand{\order}[1]{\mathcal{O}(#1)}
\newtheorem{theorem}{Theorem}
\newtheorem{lemma}[theorem]{Lemma}

\begin{document}

\begin{frontmatter}
\title{Fully Bayesian Estimation Under Informative Sampling}
\runtitle{Fully Bayes Under Informative Sampling}

\begin{aug}
\author{\fnms{Luis G.} \snm{Le\'on-Novelo}\thanksref{t2}\ead[label=e1]{Luis.G.LeonNovelo@uth.tmc.edu}}
\and
\author{\fnms{Terrance D.} \snm{Savitsky}\thanksref{t3}\ead[label=e2]{Savitsky.Terrance@bls.gov}}

\address{1200 Pressler St. Suite E805, Houston, TX, 77030, USA and \\
PSB Suite 1950, 2 Massachusetts Avenue, NE Washington, DC 20212-0001, USA.\\
\printead{e1,e2}}

\thankstext{t2}{Department of Biostatistics and Data Science,
University of Texas Health Science Center at Houston- School of Public Health}
\thankstext{t3}{Office of Survey Methods Research, Bureau of Labor Statistics}
\runauthor{Le\'on-Novelo and Savitsky}

\affiliation{Department of Biostatistics and Data Science,
University of Texas Health Science Center at Houston- School of Public Health
and Office of Survey Methods Research, Bureau of Labor Statistics}

\end{aug}

\begin{abstract}
Survey data are often collected under informative sampling designs where subject inclusion probabilities are designed to be correlated with the response variable of interest. The data modeler seeks to estimate the parameters of a population model they specify from these data.  Sampling weights constructed from marginal inclusion probabilities are typically used to form an exponentiated pseudo likelihood that adjusts the population likelihood for estimation on the sample. We propose an alternative adjustment based on a Bayes rule construction that simultaneously performs weight smoothing and estimates the population model parameters in a fully Bayesian construction. We formulate conditions on known marginal and pairwise inclusion probabilities that define a class of sampling designs where $L_{1}$ consistency of the joint posterior is guaranteed. We compare performances between the two approaches on synthetic data. We demonstrate that the credibility intervals under our fully Bayesian method achieve nominal coverage.
We apply our method to data from the National Health and Nutrition Examination Survey to explore the relationship between caffeine consumption and systolic blood pressure.
\end{abstract}

\tableofcontents
\end{frontmatter}

\section{Introduction}
Our focus is on the data analyst who seeks to conduct inference about a finite population generating model,
$P_{0}$, where the observed data are acquired through a survey sample taken from a finite population of interest.   The data analyst has a chosen population model, $P$, about which they desire to perform inference or prediction and seeks to estimate its parameters taking into account
an informative sampling design that governs the observed sample (because an informative sampling design is not ignorable as it is under simple random sampling). Our focus on population model estimation contrasts with the usual design-based estimation of simple domain statistics (\eg\ employment rate by county)
that are often of interest to statistical agencies. Of course, a well-constructed population model, $P$,  that accounts for dependencies within the observed data may also be used to construct simple domain-level statistics in an efficient manner.

Bayesian hierarchical model specifications are increasingly used to perform inference about $P_{0}$
due to their flexibility for modeling dependencies among observations and the advent of
``black box" solvers, especially ``Stan" \citep{carpenter2016stan}, which performs an
efficiently-mixing Hamiltonian Monte Carlo sampling algorithm with a feature that
non-conjugate model specifications are readily accommodated.
Bayesian population models can be flexibly parameterized and, since
they do not rely on asymptotic theory, are suitable for small sample inference.

A survey sampling procedure randomly draws the observed sample from a target finite population on which inference is performed about $P_{0}$.  A known sampling design distribution, $P_{\nu}$, defines a joint distribution over the random inclusions of units from the finite population.   The sampling design distribution often intentionally constructs a correlation between specified unit inclusion probabilities and the response variable of interest; for example, the U.S. Bureau of Labor Statistics (BLS) employs proportion-to-size (PPS) sampling design in the Current Employment Statistics (CES) survey.  The CES is administered with the purpose to construct total employment statistics for area and industry-indexed domains.  Establishments with larger employment values are assigned higher unit inclusion probabilities since these larger establishments drive more of the variance in total employment point estimates for domains.  Survey designs that instantiate a correlation between unit inclusions and the response variable of interest are termed, ``informative", because the inclusion probabilities are informed by the response variable.  The balance of information in the observed sample is different from that in the population under informative sampling, such that naively estimating a proposed population model, $P$, on the observed sample will result in biased inference about $P_{0}$, that is supposed to govern the finite population \citep{savitsky2016bayesian}.  We construct a fully Bayesian estimator in the sequel that accounts for joint distribution, $(P_{0},P_{\nu})$, over population generation and the taking of an informative sample

\subsection{Review of  Methods to Account for Informative Sampling}
There are three broad classes of approaches to adjust model estimation on the observed sample acquired under an informative sampling (IS) design for inference about $P_{0}$. One approach parameterizes the sampling design into the model estimated on the sampled data \citep{Litt:mode:2004}.  The sampling design, itself, is often a nuisance to the data analyst, however, and their focus on the parameters of $P_{0}$ requires them to marginalize over parameters indexing the sampling design distribution.  It is also often the case that the analyst does not know the sampling design to parameterize it.

The next two classes of modeling adjustments employ a sampling weight that is constructed to be inversely proportional to the marginal inclusion probability, $\pi_{i} = P\left(\delta_{i} = 1\right)$, for each unit, $i \in \left(1,\ldots,n\right)$, where $n$ denotes the number of units in the observed sample, $S$.  The inclusion of unit, $i$, from the population, $U$, in the sample, $S \subset U$ is indexed by the random variables, $\delta_{i} \in \{0,1\}$, governed by $P_{\nu}$.  The likelihood contribution for each unit in the observed sample is adjusted by its associated sampling weight, such that the joint adjusted likelihood over the sample provides an approximation to the balance of information in the finite population.

The second class of modeling approaches employs a particular form for the likelihood \citep{dong:2014, kunihama:2014, wu:2010, si2015bayesian}, rather than allowing the analyst to specify a population model, $P$, because the inferential focus is not on parameters of the generating model, $P_{0}$, but on domain-level estimation of simple mean and total statistics, such as total employment for a geographic area and industry combination.

In the third class of modeling approaches, \citet{savitsky2016bayesian} construct a sampling-weighted pseudo posterior distribution by exponentiating each unit likelihood contribution, under the analyst-specified model, by its sampling weight, to produce, $p\left(y_{i}\vert \delta_{i} = 1,\bm{\lambda}\right)^{w_{i}}$.  Exponentiating by the sampling weight, $w_{i} \propto 1/\pi_{i}$, constructs the pseudo likelihood used to estimate the pseudo posterior when convolved with the prior distributions for model parameters, $\bm{\lambda}$. \citet{savitsky2016bayesian} demonstrate that estimation of (the parameters of) $P_{0}$ from the pseudo posterior distribution is asymptotically unbiased.  This approach provides a ``plug-in" approximation to the population likelihood (for $n$ observations), in that the sampling inclusion probabilities, $(\pi_{i})$, are assumed fixed.  On the one hand, one may suppose that a sequential generation for the finite population under $P_{0}$ and the subsequent taking of the sample under $P_{\nu}$ (where the two distributions are convolved under informative sampling).  In this case, one may view the information contained in the generated finite population as fixed.  Inclusion probabilities are assigned by $P_{\nu}$ to the units comprising the population, and so, are also viewed as fixed (given the underlying finite population).  From a Bayesian perspective, on the other hand, there is a dynamic process of population generation and the taking of an informative sample.  If the population changes, then sample inclusion probabilities, which depend on the population response values, will also change.

The absence of an established procedure to incorporate the inclusion probabilities into the likelihood of a particular population model chosen by the data analyst hinders the use of Bayesian statistics in the analysis of survey data released with sampling weights.  To address this lack of a fully Bayesian treatment for modeling data under informative sampling, this paper formulates a fully Bayesian set-up to jointly model $(y_{i},\pi_{i})$ from $P$ on the observed sample.  Our approach extends the set-up of \citet{pfeffermann1998parametric} to a fully Bayesian formulation by specifying a conditional population model, $p(\pi_{i}\mid y_{i},\bka)$, for the inclusion probabilities, $(\pi_{i})_{i\in U}$.  We then apply a Bayes rule approach to construct, $p(y_{i},\pi_{i}\mid \bx_{i}, \bth, \bka,\delta_{i}=1)$, that conditions on the observed sample, $(\delta_{i} = 1)_{i = 1,\ldots, N}$, where $N$ denotes the population size, $\vert U \vert$.  We employ this full likelihood defined on the observed sample to estimate the joint posterior distribution for model parameters, $(\bm{\theta},\bm{\kappa})$.  We demonstrate in the sequel that the informativeness of the sampling design improves estimation efficiency for our Fully Bayes approach that jointly models $(y_{i},\pi_{i})$ as compared to the plug-in, pseudo posterior framework.  We formulate conditions that guarantee the $L_{1}$ frequentist consistency of our fully Bayes estimator.  We further conduct a simulation study that demonstrates the credibility sets generated under our fully Bayes construction achieve nominal coverage (and we also demonstrate, by contrast, that credibility sets of the sampling-weighted, pseudo posterior distribution do not achieve nominal coverage).

\citet{pfeffermann1998parametric} focus on maximum likelihood point estimation from $p(y_{i}\mid \mathbf{x}_{i},\bm{\theta}, \bm{\kappa},\delta_{i}=1)$, rather than the joint distribution over the response and inclusion probabilities, as do we.  We also relax their theoretical condition of independence among the sampled units that they employ to guarantee consistency of the point estimate to asymptotic independence in our result for the $L_{1}$ consistency of our fully Bayes posterior distribution.

\citet{Pfef:DaS:DoN:mult:2006} also extend \citet{pfeffermann1998parametric} to Bayesian estimation, but they employ $p(y_{i}\mid \bx_i, \bth, \bka,\delta_{i}=1)$, (without specifying a model for the $(\pi_{i})$) rather than our formulation for the joint likelihood, which perhaps conforms to the view that the inclusion probabilities, $(\pi_{i})$, are \emph{not} generated along with the population, $(y_{i},\mathbf{x}_{i})_{i=1,\ldots,N},~N=\vert U \vert$, but are fixed under specification of the sampling design distribution, $P_{\nu}$.  So their approach may be viewed to be \emph{not} fully Bayesian as is ours because it does not specify a model for $p(\pi_{i}|y_{i},\bka)$.

We derive the joint likelihood for the observed sample, $p(y_{i},\pi_{i}\vert \delta_{i} = 1,\cdots)$, using a Bayes rule approach that adjusts the analyst-specified population likelihood, $p(y_{i},\pi_{i}\vert \cdots)$, from which we formulate our joint posterior distribution for the observed sample in Section~\ref{method}.   Conditions are constructed that guarantee a frequentist $L_{1}$ contraction of our fully Bayesian posterior distribution, $P$, onto the true generating distribution in Section~\ref{theory}. This section can be skipped without loss of understanding the practical application of the proposed method. We conduct a simulation study under a synthetically generated population from a known, $P_{0}$, from which we draw samples under an informative, proportion-to-size (PPS) sampling design, to compare the performances of our fully Bayesian estimator and the pseudo posterior in Section~\ref{sec:simulation}.
We reveal that: (1) fully Bayes credible intervals achieve nominal frequentist coverage
while the pseudo posterior under covers; and (2) the fully Bayes point estimates are robust, in terms of bias and mean square error (MSE), against high variability of inclusion probabilities in contrast to the pseudo posterior.
We illustrate our fully Bayesian estimator by assessing the relationship between caffeine consumption and systolic blood pressure using NHANES data in Section~\ref{sec:application}.
The paper concludes with a discussion in Section~\ref{sec:discussion}.
Supplementary Material is provided in  Sections \ref{sec:appendixprooftheoremcloseform}-\ref{sub:datapreprocess}. This includes the proofs for the main result, along with two enabling results in  Section \ref{subsec:suppleLinReglikelihood}.

\section{Fully Bayesian Estimator under Bayes Rule}\label{method}
We proceed to formulate a posterior distribution for the \emph{observed} sample taken under an informative design as a function of inclusion probabilities to produce unbiased inference and correct uncertainty quantification for the population model.  Our approach first constructs a joint distribution for the response and inclusion probabilities for the \emph{population} in a fully Bayesian specification.  We then use Bayes rule to adjust the joint population likelihood to an expression for the observed sample by incorporating conditional distributions for inclusion indicators, $(\delta_{i} \in\{0,1\})$.  The adjusted likelihood is unbiased for the population model parameters estimated on the sample.  We later show that credibility sets of the posterior distribution estimated from the fully Bayes adjusted likelihood achieve nominal coverage.

\noindent From Bayes rule,

\begin{align}
  p(y_{i},\pi_{i}\vert \bx_{i}, \bth, \bka,\delta_{i}=1)=\frac
{\mbox{Pr}(\delta_i=1 \vert y_{i},\pi_{i},\bx_{i}, \bth, \bka  )\times p(y_i,\pi_i\vert \bx_i, \bth, \bka)   }
{\mbox{Pr}(\delta_i=1 \vert \bx_i, \bth, \bka)}\label{eq:samplingdist},
\end{align}
where the expression on the left-hand side of Equation~\eqref{eq:samplingdist} adjusts the joint population likelihood, $p(y_i,\pi_i\vert \bx_i, \bth, \bka)$ , specified on the right-hand side for the realized sample.

\noindent We may simplify the numerator by plugging in \citep[See also equation (7.1) in][]{pfeffermann1998parametric} ,
\begin{equation}\label{eq:deltagivenall}
{\mbox{Pr}}(\delta_i=1 \vert y_{i},\pi_{i},\bx_{i}, \bth, \bka  ) = \pi_i.
\end{equation}

\noindent We next compute the double expectation with respect to $(\bka, \bth)$ in the denominator,
   \begin{align}
    \mbox{Pr}(\delta_i=1 \vert \bx_i, \bth, \bka)=&
    \int \int \mbox{Pr}(\delta_i=1 \vert y_i,\pi_i,\bx_i, \bth, \bka)  p( y_i,\pi_i\vert \bx_i, \bth, \bka)\,d\pi_i d y_i\label{eq:marginaldelta}\\
    =&\int \int   \pi_i    p( \pi_i\vert y_i, \bx_i,\bka)\, d\pi_i\, p( y_i\vert \bx_i, \bth) \,d y_i\nonumber\\
=&\int E(\pi_i\vert y_i, \bx_i, \bka)\, p( y_i\vert \bx_i, \bth) \,d y_i\nonumber\\
  =&E_{y_i\vert \bx_i,\bth}\left[E(\pi_i\vert y_i, \bx_i, \bka) \right]\nonumber,
          \end{align}

\noindent where we assume separability of the parameter sets, $\bka, \bth$, in the conditional distributions for $\pi_i\vert y_i,\bx_i$ and $y_i\vert \bx_i$,
\begin{align}\label{eq:jointyandpi}
    p(y_i,\pi_i\vert \bx_i, \bth, \bka)=&p(\pi_i\vert y_i,\bx_i,\bth, \bka) p(y_i\vert \bx_i,\bth, \bka)\\
             =&  p(\pi_i\vert y_i,\bx_i,\bka) p(y_i\vert \bx_i,\bth),\nonumber
\end{align}
where $p(\pi_i\vert y_i,\bx_i,\bka)$ and $p(y_i\vert \bx_i,\bth)$ denote conditional distributions for the population.  We make note that we have factorized the joint population distribution over $(y_{i},\pi_{i})$, such that information about the sampling design is parameterized in the conditional model for $\pi_{i}\vert y_{i}$, rather than in the conditional distribution for $y_{i}\vert x_{i}$.

We define $p_s(y_{i},\pi_{i}\vert \bx_{i}, \bth, \bka)=p(y_{i},\pi_{i}\vert \bx_{i}, \bth, \bka,\delta_{i}=1)$  (where subscript, $s$, indexes the observed sample of size, $n$) to be the likelihood contribution for each unit, $i \in S$, and
plug in Equations~ \eqref{eq:deltagivenall} and \eqref{eq:marginaldelta} into Equation~\eqref{eq:samplingdist} to obtain,
\begin{equation}\label{eq:IScorrection}
p_s(y_{i},\pi_{i}\vert \bx_{i}, \bth, \bka)=\frac
{\pi_i   p(\pi_i\vert y_i,\bx_i,\bka)   }
{E_{y_i^\prime\vert \bx_i,\bth}\left[E(\pi_i^\prime\vert y_i^\prime, \bx_i,  \bka) \right]}
\times p( y_i\vert \bx_i, \bth)
\end{equation}

Let $N = \vert U\vert$ denote the population size and $S=\{S(1),\dots,S(n)\}:=\{i\in \{1,\dots,N\}:\delta_i=1\}$, the population indices of individuals or units included in the random sample $S$ of size $n = \vert S \vert$. Therefore, ${\mbox{Pr}}(S=s)\propto \prod_{i=1}^{n} \pi_{s(i)}$.
Let the generic vector $\sampled{r}:=\{\is{r_1},\dots,\is{r_n}\}:=\{r_{s(1)},\dots,r_{s(n)}\}$
represent the values of variable, $r$, observed in the sample.
With this notation the likelihood over the observed sample of size, $n$, is specified by,
 \begin{equation*}\label{eq:likelihood}
\ell(\bth,\bka;\sampled{y},\sampled{\pi},\sampled{x})=\prod_{i=1}^n\left[p_s(\is{y_i},\is{\pi}_i\mid \sampled{x_i},\bth,\bka) \right]
\end{equation*}
from which we may incorporate the prior distributions to formulate the posterior distribution observed on the sample,
$$
p_s(\bth,\bka\mid\sampled{y},\sampled{\pi},\sampled{x})\propto \
\ell(\bth,\bka;\sampled{y},\sampled{\pi},\sampled{x})
\times \hbox{Prior}(\bth)\times \hbox{Prior}(\bka).
$$

We have now constructed the backbone of a fully Bayesian model that takes into account informative sampling designs for estimation of population model parameters.  Our result jointly models the response and the inclusion probabilities, \ie, $(y_i,\pi_i)$, using only quantities observed in the sample; in particular, the joint distribution of $(y_i,\pi_i)$ are different in the observed sample and in the population, and we have corrected for this difference in a way that allows us to make unbiased estimation of the parameters of the population model.
In contrast to \cite{si2015bayesian}, we do not impute the values for the non sampled units.
By assuming a population distribution for the inclusion probabilities,
the variability of the inclusion probabilities is modeled and the noise
not depending on the response (and $\bx_i$)
is discarded; more specifically $p_s(y_i,\pi_i\mid \cdots)$ in Equation~\eqref{eq:IScorrection}
is constructed from a conditional model that regresses $\pi_{i}$ on $y_{i}$ and uses the expected value in the denominator.

Achieving our result of Equation~\eqref{eq:IScorrection} requires specification of a joint population generating distribution, $p(y_i,\pi_i\vert \bx_i, \bth, \bka)$.
Specifying a population model for the response, $p(y_i\vert \bx_i,\bth)$, is commonly done.
Going a step further, as we do, to incorporate the inclusion probabilities in the population model, is not commonly done. By contrast, the pseudo posterior approach, which is formally introduced in the sequel,
considers the inclusions probabilities (or equivalently, the sampling weights) as fixed.
Under an informative sampling design where the $(\pi_{i})$ are constructed to depend on the $(y_{i})$, each time we generate new values of the response variable we are prompted to update the inclusion probabilities for all units in the population after fixing a sampling design.  So we believe the treatment of $\pi_{i}$ as random is natural as they are conditioned on the values of population variables.

The price the modeler pays for this fully Bayesian approach is that they are required to
specify a conditional distribution of the inclusion probabilities for all units in the population,
$p(\pi_i\mid y_i,\bka)$. This requirement raises two computational difficulties:
Firstly, even if the specified population likelihood,
$p(y_i\vert \bx_i, \bth)$, and the associated prior, $p\left(\bth\right)$,  yield a closed form conditional posterior distribution, this construction for the population model does not produce a conjugate posterior under our formulation to correct for informative sampling (IS) because the sampling design is not ignorable; Secondly, the computation of the expected value in the denominator of  Equation~\eqref{eq:IScorrection} is
a computational bottleneck as we are required to compute
this expected value for each observation in every Gibbs sampler iteration.
We cope with the first difficulty by relying on the black box solver Stan~\citep{carpenter2016stan}, that uses a Hamiltonian Monte Carlo approach to draw samples from the full joint posterior distribution, so that it performs well under (is insensitive to) non-conjugacy. Regarding the second computational challenge, we proceed
to construct a set of conditions on $p(\pi_i\mid y_i,\bx_i,\bka)$ and the likelihood that guarantee the availability of a closed form expression
for this expected value.

We next specify a class of conditional population distributions from Equation~\eqref{eq:jointyandpi}
that yield a closed form result for the expectation step in denominator of \eqref{eq:IScorrection}, which simplifies posterior computation.
Let $\bxp_i$ and $\bxy_i$ be subvectors of $\bx_i$, the covariates used to specify the conditional distribution of $\pi_i\mid y_i,\bx_i,\bka$
and $y_i\mid \bx_i,\bth$, respectively; that is,
$\pi_i\mid y_i,\bx_i,\bka \sim \pi_i \mid y_i,\bxp_i,\bka$ and
 $y_i\mid \bx_i,\bth\sim y_i\mid \bxy_i,\bth$.  Note that we allow for $\bxp_i$ and $\bxy_i$ to have common covariates.
Let $\hbox{normal}(x\mid\mu,s^2)$ denote the normal distribution pdf with mean $\mu$ and variance $s^2$ evaluated at $x$, and $\hbox{lognormal}(\cdot\mid\mu,s^2)$ denote the lognormal pdf, so that
$X\sim     \hbox{lognormal}(\mu,s^2)$ is equivalent to  $\log X\sim\hbox{normal}(\mu,s^2)$.
Going forward, we assume that $\pi_i$ is proportional, as opposed to exactly equal,
to the inclusion probability for unit $i$. In other words, no restriction is imposed on $\sum_{i} \pi_i$ where
the index $i$ could run over the population or sample indices.  We will see in the sequel that normalizing
the sum of the inverse inclusion probabilities \emph{is} required to regulate the estimation of posterior uncertainty under the pseudo likelihood approach, but such is not required for our fully Bayesian formulation.

\begin{theorem}\label{th:closeform}
If
$p(\pi_i\mid y_i,\bxp_i,\bka) =\emph{lognormal}(\pi_i\mid h(y_i,\bxp_i,\bka),\sigma_{\pi^2})$,
with the function $h(y_i,\bxp_i,\bka)$ of the form $h(y_i,\bxp_i,\bka)=\hy(y_i,\bxp_i,\bka)+\hmy(\bxp_i,\bka)$ where
$\sigma_{\pi}^2=\sigma_{\pi}^2(\bka,\bxp_i)$, possibly a function of $(\bka,\bxp_i)$
then
$$
p_s(y_i,\pi_i\mid \bxy_i,\bxp_i,\bth,\bka)=
\frac{\emph{normal}\left(\log \pi_i\mid \hy(y_i,\bxp_i,\bka)+\hmy(\bxp_i,\bka),\sigma_\pi^2\right)}
       {\exp\left\{\hmy(\bxp_i,\bka)+\sigma^2_\pi/2\right\}  \times M_y(\bka;\bxy_i,\bxp_i,\bth)    }
\times p(y_i\mid \bxy_i,\bth)\nonumber
$$
 with $M_y(\bka;\bxy_i,\bxp_i,\bth):=E_{y\mid \bxy_i,\bth}\left[\exp\left\{\hy(y_i,\bxp_i,\bka)\right\}\right]$.
\end{theorem}

The proof of this theorem is in the Supplementary Material, Section~ \ref{sec:appendixprooftheoremcloseform}.
If both $M_y$ and $p(y_i\mid\cdots)$ admit closed form expressions, then $p_s(y_i,\pi_i\mid\cdots)$ has a closed form, as well;
for example, when 
$\hy(y_i,\bxp_i,\bka)=\kappa_y y_i$
with $\kappa_y\in \bka \in \mathbb{R}$,
then $M_y(\bka;\bxy_i,\bxp_i,\bth)$ is the moment generating function (MGF) of $y_i\mid \bth$ evaluated at $\kappa_y$, which will have a closed form defined on $\mathbb{R}$ for typically-used normal and binomial population models for the responses. The closed form for $M_y(\bka;\bxy_i,\bxp_i,\bth)$ implies a closed form for
$p_s(y_i,\pi_i\mid\cdots)$.
Analogously, we may consider an interaction between $y$ and $\bxp$, using
$\hy(y_i,\bxp_i,\bka)=(\kappa_y+\bxp_i^t \bka_\bxp) y_i\equiv t y_i$ with
$\bka=(\kappa_y,\bka_\bxp)$. In this case, we achieve,  $M_y(t;\cdots)$, which is the MGF evaluated at $t$.
Our assumption of a lognormal distribution for $\pi_{i}$ is mathematically appealing since the inclusion probability, $\pi_i$, for individual, $i$, is composed from the product of inclusion probabilities of selection across the
stages of the multistage survey design. If each of these stagewise probabilities are {lognormal} then their product, $\pi_i$, is {lognormal} as well. We next apply Theorem~\ref{th:closeform} to some common settings.

\subsection{Linear Regression Population Model}\label{subsec:LRM} Assume the linear regression model for the population,
$p(y_i\mid \bxy_i,\bth)$, is constructed as,
 \begin{equation}\label{eq:SLR_likelihood}
{y_i\mid \bxy_i,\bth}\sim\text{normal}\left(\bxy_i^t\bbe ,\sigma_y^2 \right) , \quad\hbox{with }\bth=(\bbe,\sigma_y^2)
\end{equation}
and the conditional population model for inclusion probabilities is specified as,
\begin{equation}\label{eq:lonnormalpriorforpi}
{\pi_i\mid y_i,\bxp_i,\bka}\sim
\text{lognormal}\Big(\kappa_y y_i+\bxp_i^t \bka_x ,   \sigma_\pi^2\Big),\quad\hbox{with }
                                                                                   \bka=(\kappa_y,\bka_x,\sigma_\pi^2)
\end{equation}
This construction results from setting, $\hy(y_i,\bxp_i,\bka)=k_y y_i$,
$\hmy(\bxp_i,\bka)=\bxp_i^t \bka_x$,
$\sigma_\pi^2(\bka,\bxp_i)=\sigma_\pi^2$.
 Here $\bbe$ and $\bka_x$ are vectors of regression coefficients
 that include an intercept, so the first entry of both $\bxy_i$ and $\bxp_i$ equals 1.
 We select prior distributions,
 \begin{align}\label{eq:priors}
 \bbe\sim& \hbox{MVN}(\mathbf{0},100 \mathbf{I})\nonumber\\
 \bka \sim& \hbox{MVN}(\mathbf{0},100 \mathbf{I})\\
 \sigma_y^2,\sigma_\pi^2 \iid& \hbox{normal}^+(0,1)\nonumber
 \end{align}
 with $\hbox{normal}^+(m,s^2)$
 denoting a normal distribution with mean $m$ and variance $s^2$ restricted to the positive real line;
 $\hbox{MVN}(\mathbf{m},\Sigma)$ denotes the multivariate normal distribution with
 mean vector $\mathbf{m}$ and variance-covariance matrix $\Sigma$; and
 $\mathbf{I}$ the identity matrix.

Since
$y\sim\hbox{normal}(m,s^2)$ admits a closed form expression for $M_y(t)=\exp(tm+t^2 s^2/2)$,
we apply Theorem \ref{th:closeform} to obtain,
\begin{align}\label{eq:LRp_s}
p_s(y_i,\pi_i\mid \bxy_i,\bxp_i,\bth,\bka)
=&\frac{\hbox{normal}(\log \pi\mid \kappa_y y_i+\bxp_i^t\bka_x    ,\sigma_\pi^2)}
       {\exp\left\{\bxp^t_i \bka_x +\sigma^2_\pi/2+  \kappa_y \bxy^t_i\bbe+\kappa_y^2\sigma_y^2/2     \right\}}\\
&\times \hbox{normal}(y_i\mid \bxy^t_i\bbe ,\sigma^2_y)\nonumber
\end{align}

The resulting form of the expression for $\log p_s(\dots)$  is provided in the Supplementary Material, Section~
\ref{subsec:suppleLinReglikelihood}.

\subsection{Other Population Models}
Since a closed-form moment generating function of
$y_i\mid \bxy_i,\bth$  defined in $\mathbb{R}$ implies
a closed form for $p_s(y_i,\pi_i\mid\dots)$,
extensions for the logistic, probit and Poisson regressions to perform asymptotically unbiased estimations on the observed sample are straightforward; for example,  employing the conditional lognormal model, Equation~\eqref{eq:lonnormalpriorforpi}, for the inclusion probabilities, but now selecting a probit construction for a dichotomous, $y_i\sim \hbox{Bernoulli}(p_i)$, with $p_i=\Phi^{-1}(\bx_i^t \bth)$, where
$\Phi^{-1}$ is the inverse of the
standard normal CDF. We note the closed form MGF, $E(e^{ty})=(1-p+pe^t)$ for all $t$,
and obtain,
\begin{align*}
p_s(y_i,\pi_i\mid \bxy_i,\bxp_i,\bth,\bka)
=&\frac{\hbox{normal}(\log \pi_i\mid \kappa_y y_i+\bxp_i^t\bka_x    ,\sigma_\pi^2)}
       {\exp\left\{\bxp^t_i \bka_x +\sigma^2_\pi/2\right\}  (1-p_i+p_ie^{\kappa_y}) }
\times \hbox{Bernoulli}(y\mid p_i)\nonumber
\end{align*}
with $\hbox{Bernoulli}(y\mid p_i)=p_i^{y_i}(1-p_i)^{1-y_i}$.

\subsection{Splines Basis Population Model}\label{subsection:spline}
We extend the linear regression model to a splines setting for situations
where the relationship between the response and the predictor is not linear.
Here, $p(y_i\mid \bxy_i,\bth)$ is constructed as,
\begin{equation}\label{eq:Splinemodel}
y_i\sim \hbox{normal}(\mu(\bxy_i),\sigma_y^2)\quad\hbox{with }
\mu(\bxy_i):=B(\bxy_i)\bbe\end{equation}
with $B(\bxy_i):=(B_{i1},\dots,B_{ib})$, the vector of penalized B-spline coefficients associated with $\bxy_i$, where $b$ denotes the number of spline bases.  We construct a rank-deficient multivariate Gaussian prior for the coefficients, $\bbe$, which penalizes or regulates the smoothness of the estimated function, $\mu$, by employing a $b\times b$ matrix, $Q=D^t D$, where $D$ is the discretized $k$th difference operator \citep{speckman2003fully} in the density kernel,
\begin{equation}\nonumber
\sigma_{\bbe}^{-(b-k)/2}   \exp{-\frac{1}{2 \sigma_{\bbe}^2} \bbe^t Q \bbe}.
\end{equation}
The order, $k$, regulates the resulting smoothness of the fitted function or its order of differentiability.  $B(\bxy_i)$ has at most $k$ entries different from zero.  The penalized pdf specified for $\bbe$ is not proper since the dimension of $\bbe$ is $b$, but $Q$ is of rank $b-k$. We observe that the MLE for $(\bbe,\sigma^2_y,\sigma^2_{\bbe})$
is a type of ridge regression estimator since it satisfies,
$$
(B^tB/\hat{\sigma}^2_y+ Q/\hat{\sigma}_{\bbe}^2)\hat{\bbe}=B^t\by/\hat{\sigma}_y^2.
$$
\noindent Our simulation scenarios in Section~\ref{sec:simulation} will utilize this penalized B-spline formulation with $(b=8,k=4)$.
The priors for model variances are set as weakly informative proper priors,
$$
\sigma_y\sim \text{normal}^+(0,10), \quad
\sigma_{\bbe}\sim  \text{Cauchy}^+(0,10) \quad \hbox{and }        \sigma_\pi\sim \hbox{Cauchy}(0,1);
$$
where $\text{Cauchy}^+(m,v)$ denotes the Cauchy distribution with location and scale parameters $m$ and $v$, respectively, restricted to be positive.
$\tau_{\bbe}=1/\sigma^2_{\bbe}$ is referred to as a complexity parameter.

Since the spline model is a particular case of the linear regression model
in Subsection \ref{subsec:LRM},
$p_s(y_i,\pi_i\mid \bxy_i,\bxp_i,\bth,\bka)$, with
$\bth:=(\bbe,\sigma_y^2)$,
for the spline model has the same expression as that for the linear model in Equation~\eqref{eq:LRp_s} after replacing $\bxy_i^T\bbe$ for $B_i\bbe$ (assuming our continued employment of a lognormal prior for $\pi_i$ in Equation~\eqref{eq:lonnormalpriorforpi}).

\subsection{Estimation of Population Distribution of Inclusion Probabilities}
For a simple one-stage design the whole population of inclusion probabilities or equivalent is available, because their values are required to draw a sample. For a multistage cluster sample like a survey of individuals within households, we draw a sample recursively, only specifying inclusion probabilities for units selected at the previous level.  This procedure provides sampling probabilities only for those lowest-level units or individuals that are selected into the sample.  It may be useful for the survey administrator to estimate the population distribution of inclusion probabilities from the observed sample in order to verify a subgroup of units targeted for over-sampling do, indeed, express relatively higher inclusion probabilities.

\begin{figure}
\begin{center}
\includegraphics [width=70mm,angle=0]{\plotpath{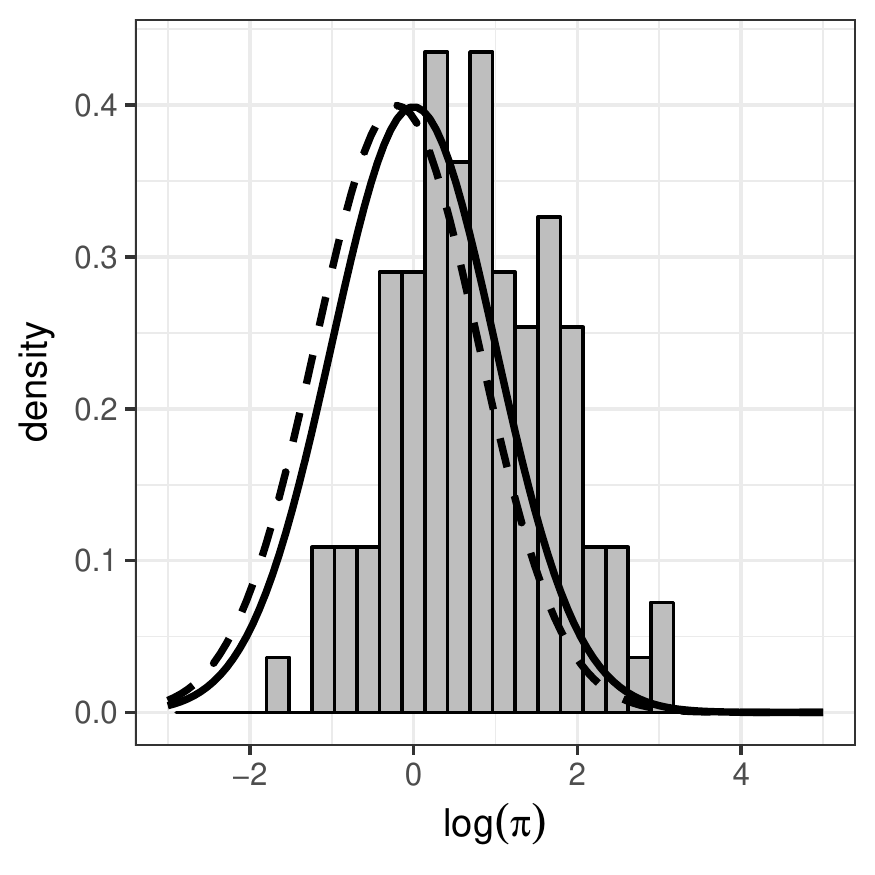}}\\
\end{center}
\vskip-5mm
\caption{\label{fig:weightshistogram}
Histogram of the $n=100$ $\log$-transformed \emph{sampled} inclusion probabilities
along with the true simulated standardized normal population density
(solid line) and estimated normalized population density (dashed line) with mean and variance estimated using Equation~\eqref{eq:piunderIS}.
}
\end{figure}

We select a model that satisfies the conditions of Theorem \ref{th:closeform},
but set $y_i\equiv 0$. This is,  $\hy(y_i,\bxp_i,\bka)$ is constant and without loss of generality
equal to zero. Then, $M_y(\bka;\bxy_i,\bxp_i,\bth)=1$ and Theorem \ref{th:closeform} implies
\begin{equation}\label{eq:piunderIS}
p_s(\pi_i\mid \bxp_i,\bka)=
\frac{\hbox{normal}\left(\log \pi_i\mid \hmy(\bxp_i,\bka),\sigma_\pi^2\right)}
       {\exp\left\{\hmy(\bxp_i,\bka)+\sigma^2_\pi/2\right\}   }
\end{equation}
If we assume $\pi_i\sim \hbox{lognormal}(\hmy(\bxp_i,\bka),\sigma_\pi^2)$ for $i=1,\dots,N$ with
$\mbox{Pr}(\delta_i = 1)= \pi_i/\sum_{i=1}^N \pi_i$,
we can make inference about the parameters $\bka$ and $\sigma_\pi^2$ using only the observed inclusion probabilities.
As an example we simulate $\pi_1,\dots,\pi_{10^5}\iid \hbox{lognormal}(\bka=0,\sigma_\pi^2=1)$. (Our choice of the lognormal distribution implies $\hmy(\bxp_i,\bka)\equiv\bka$.)  We draw $n=100$ inclusion probabilities $(\pi_i)$ under IS, and estimate the parameter values $\bka$ and $\sigma_\pi^2$ via Equation~\eqref{eq:piunderIS} using the priors outlined in Equation~\eqref{eq:priors}.
Figure \ref{fig:weightshistogram} shows the histogram of the log transformed sampled inclusion probabilities observed in our sample of $n=100$, along with the standardized density of generated inclusion probabilities for our simulated population, displayed in the solid line, and our estimated normal density with population
mean $E(\bka\mid \pi_1,\dots,\pi_n)$ and variance
$E(\sigma_\pi^2\mid \pi_1,\dots,\pi_n)$, displayed in the dashed line.
Notice that we employ the unstandardized sampling
weights. If the sampling weights are multiplied by a constant $c > 0$;
e.g., $c:=n/\sum_{i=1}^n \pi_i$, we are then estimating $\bka+\log c$.

\subsection{Pseudolikelihood Alternative to the Fully Bayes Estimator}
The pseudo posterior approach is a plug-in method that is not fully Bayesian, which
we will use to compare with the fully Bayes method we propose in this paper.
The pseudo likelihood is formed by exponentiating the likelihood contribution for each observed unit in the sample by its sampling weight, $\is{w_i}$, in,
$$
\mbox{pseudo}(\bth; \sampled{y},\sampled{w})=\prod_{i=1}^n p(\is{y_i}\mid \sampled{x}_i,\bth)^{\is{w_i}}
$$
where $\is{w_i}\propto 1/\is{\pi_i}$ is the sampling weight that is standardized such that $\sum \is{w_i}=n$, where $n$ denotes to the observed sample size, which is required to regulate the amount of estimated posterior uncertainty.
Let $\sampled{w}:=(\is{w_1},\dots,\is{w_n})$ denote the vector of these unit-indexed sampling weights that are typically published with the survey data in order to
correct for IS. Inference is carried out by constructing the pseudo posterior distribution be proportional to
$\mbox{pseudo}(\bth; \sampled{y},\sampled{w})~\times~ \mbox{Prior}(\bth)$
\citep{savitsky2016bayesian}.  Although the pseudo likelihood is an improper distribution, it admits a proper joint posterior distribution under employment of proper priors.  Classical implementations of the sampling weighted pseudo likelihood, by contrast to the Bayesian implementation, employ MLE estimation of population model parameters on the observed IS. The standard error of the estimate of parameter, $\theta$, is estimated via resampling methods such as
balanced repeated replication \citep{mccarthy1969pseudo,krewski1981inference} or Fray's method \citep{judkins1990fay} when the sampling design and the information provided to the analyst allow it.

The advantages of the plug-in pseudo posterior method over the proposed fully Bayesian construction are:
$(i)$ Estimation does not require a custom posterior sampler. Relatively slight modifications are performed to the population model sampler to incorporate the pseudo likelihood;
$(ii)$ Specification of $\pi_i\mid y_i,\cdots$ for the population is not required;
$(iii)$ There is no expected value in Equation~\eqref{eq:marginaldelta} to compute as there is in the fully Bayes method.
The disadvantages of the pseudo posterior approach are:
$(i)$ It is not fully Bayesian; $(ii)$ It does not discard variation in weights that do not depend on the response, such that noise (unrelated to the response) is introduced into the estimation of the pseudo posterior distribution; $(iii)$ The weights must be normalized to regulate the amount of estimated posterior uncertainty, which is not required for the fully Bayes approach.  We will show in the sequel that the fully Bayes approach expresses superior performance in estimation of posterior uncertainty than does the pseudo posterior.

\section{Posterior Consistency}\label{theory}
We proceed to construct conditions on the population generating model and sample inclusion probabilities that guarantee the contraction of our fully Bayes estimator to the true joint posterior distribution.  The conditions constructed for the sample inclusion probabilities define a class of sampling designs for which we would expect asymptotically unbiased estimation of our population model parameters on the observed data.
\subsection{Set-up}
Let $\nu \in \mathbb{Z}^{+}$ index a sequence of finite populations, $\{U_{\nu}\}_{\nu=1,\ldots,N_{\nu}}$, each of size, $\vert U_{\nu} \vert = N_{\nu}$, such that $N_{\nu} < N_{\nu^{'}},\text{ for}~\nu < \nu^{'}$, so that the finite population size grows as $\nu$ increases.  We employ subscript, $\nu$, because the process of rendering a sample involves the generation of a population followed by the assignment of sample inclusion probabilities and the taking of a sample from that population using the sampling weights.  The entire sequence is repeated on each increment of $\nu$. A sampling design is defined by placing a distribution on a vector of inclusion indicators, $\bm{\delta}_{\nu} = \left(\delta_{\nu 1},\ldots,\delta_{\nu N_{\nu}}\right)$, linked to the units comprising the population, $U_{\nu}$, that we use to draw a sample of size $n_{\nu} < N_{\nu}$.  We construct a sampling distribution by specifying marginal inclusion probabilities, $\pi_{\nu i} = \mbox{Pr}\{\delta_{\nu i} = 1\vert Y_{\nu,i} = y_{\nu i}\}$ for all $i \in U_{\nu}$ and the second-order pairwise probabilities, $\pi_{\nu ij} = \mbox{Pr}\{\delta_{\nu i} = 1 \cap \delta_{\nu j} = 1\vert Y_{\nu,i} = y_{\nu i}, Y_{\nu,j} = y_{\nu j}\}$ for $i,j \in U_{\nu}$, where $Y_{\nu,1},\ldots, Y_{\nu,N_{\nu}}$ index random variables that are independently distributed according to some unknown distribution $P_{\theta}$ (with density, $p_{\theta}$) defined on the sample space, $\left(\mathcal{Y}_{\nu},\mathcal{A}_{\nu}\right)$.

In the usual survey sampling set-up, the inclusion probabilities are treated as \emph{fixed}.  Under the Bayesian paradigm, however, $Y_{\nu}$ is random variable drawn from a super-population, such that if a new population, $U_{\nu}$, with associated observed values $(y_{\nu i})$, is drawn, the marginal inclusion probabilities, $(\pi_{\nu i})$, will update.   So we condition the inclusion probabilities explicitly on the population response values for emphasis, unlike in \citet{savitsky2016bayesian}.  We envision both inclusion probabilities, $(\pi_{\nu i})$, and our data, $(y_{\nu,i})$, as generated from a super-population under some true model.  We construct the joint density of the model for $(y_{\nu i},\pi_{\nu i})$ for each $i = 1,\ldots,N_{\nu}$ as,
\begin{equation}
p_{\lambda}\left(y_{\nu i},\pi_{\nu i}\right) = p_{\theta}\left(y_{\nu i}\right)\times p_{\kappa}\left(\pi_{\nu i}\vert Y_{\nu i}=y_{\nu i}\right),
\end{equation}
where $\lambda = (\theta,\kappa)$ denotes our model parameters.

We use the Bayes rule construction we earlier illustrated to formulate a sampling design-adjusted fully Bayesian joint likelihood for the unknown population value, $\lambda_{0}$, which we suppose generates the finite population values that we observe in our sample, $\left(y_{\nu,i},\pi_{\nu,i}\right)_{i\in U_{\nu}: \delta_{\nu i} = 1}$,
\begin{align}
p_{\lambda}^{\pi}\left(x_{\nu i}\delta_{\nu i}\right)& \\
&:= \left[\frac{\pi_{\nu i} \times p_{\lambda}\left(y_{\nu i},\pi_{\nu i}\right)}{\mathbb{E}_{\theta}\left\{\mbox{Pr}_{\kappa}\left(\delta_{\nu i} = 1\vert Y_{\nu i} = y_{\nu i}\right)\right\}}\right]^{\delta_{\nu i}}\\
&:= \left[\frac{\pi_{\nu_i}}{\mathbb{E}_{\theta}\left(\pi_{\nu i}^{\kappa}\right)}\times p_{\lambda}\left(y_{\nu i},\pi_{\nu i}\right)\right]^{\delta_{\nu i}}\\
&:= \left[\frac{\pi_{\nu_i}}{\pi_{\nu i}^{\lambda}}~p_{\lambda}\left(y_{\nu i},\pi_{\nu i}\right)\right]^{\delta_{\nu i}},~i \in U_{\nu}\label{bayesrule},
\end{align}
where we collect, $x_{\nu i} = \left(y_{\nu i},\pi_{\nu i}\right)$,
\newline
and define $\pi_{\nu i}^{\kappa} := \mbox{Pr}_{\kappa}\left(\delta_{\nu i} = 1\vert Y_{\nu i} = y_{\nu i}\right) =
\mathbb{E}_{\kappa}\left(\pi_{\nu i} \vert Y_{\nu i} = y_{\nu i}\right)$, where the second equality is derived in
\citet{pfeffermann1998parametric}. We further define $\pi_{\nu i}^{\lambda} := \mathbb{E}_{\theta}\left(\pi_{\nu i}^{\kappa}\right) = \mathbb{E}_{\theta}\left\{\mathbb{E}_{\kappa}\left(\pi_{\nu i} \vert Y_{\nu i} = y_{\nu i}\right)\right\}$ for ease-of-reading.

We conduct Bayesian inference under the likelihood of Equation~\eqref{bayesrule} by assigning prior, $\Pi$, on the parameter space, $\Lambda$, such that $\lambda_{0} \in \Lambda$, which produces the sampling design-adjusted posterior mass,
\begin{equation}\label{inform_post}
\Pi^{\pi}\left(B\vert x_{\nu 1}\delta_{\nu 1},\ldots,x_{\nu N_{\nu}}\delta_{\nu N_{\nu}}\right) = \frac{\mathop{\int}_{\lambda \in B}\mathop{\prod}_{i=1}^{N_{\nu}}\frac{p_{\lambda}^{\pi}}{p_{\lambda_{0}}^{\pi}}(x_{\nu i}\delta_{\nu i})d\Pi(\lambda)}{\mathop{\int}_{\lambda \in \Lambda}\mathop{\prod}_{i=1}^{N_{\nu}}\frac{p_{\lambda}^{\pi}}{p_{\lambda_{0}}^{\pi}}(x_{\nu i}\delta_{\nu i})d\Pi(\lambda)},
\end{equation}
that we use to formulate our theoretical result.  We define $\displaystyle p_{\lambda_{0}}^{\pi}(x_{\nu i}\delta_{\nu i}) = p_{\lambda_{0}}^{\pi}(x_{\nu i})^{\delta_{\nu i}}$, that confines evaluation to the observed sample.

We will utilize the empirical distribution construction to establish a bound used to prove our result,
\begin{equation}
\mathbb{P}^{\pi}_{N_{\nu}} = \frac{1}{N_{v}}\mathop{\sum}_{i=1}^{N_{\nu}}\frac{\delta_{\nu i}}{\pi_{\nu i}^{\lambda}}\delta\left(X_{i}\right),
\end{equation}
which is nearly identical to \citet{savitsky2016bayesian}, only now the denominator is a modeled quantity dependent on $\lambda$.  We construct the associated expectation functional, $\mathbb{P}^{\pi}_{N_{\nu}}f = \frac{1}{N_{v}}\mathop{\sum}_{i=1}^{N_{\nu}}\frac{\delta_{\nu i}}{\pi_{\nu i}^{\lambda}}f\left(X_{i}\right)$.  We make a similar adjustment to the Hellinger distance,
\newline
$d^{\pi,2}_{N_{\nu}}\left(p_{\lambda_{1}},p_{\lambda_{2}}\right) = \frac{1}{N_{\nu}}\mathop{\sum}_{i=1}^{N_{\nu}}\frac{\delta_{\nu i}}{\pi_{\nu i}^{\lambda}}d^{2}\left(p_{\lambda_{1}}(x_{\nu i}),p_{\lambda_{2}}(x_{\nu i})\right)$,
\newline
where $d\left(p_{\lambda_{1}},p_{\lambda_{2}}\right) = \left[\mathop{\int}\left(\sqrt{p_{\lambda_{1}}}-\sqrt{p_{\lambda_{2}}}\right)^{2}d\mu\right]^{\frac{1}{2}}$ (for dominating measure, $\mu$).

\subsection{Main Result}\label{results}
We outline the $6$ conditions and our result for completeness, though they are very similar to \citet{savitsky2016bayesian} save for the replacement of $\pi_{\nu i}$ with the modeled $\pi_{\nu i}^{\lambda}$.  The first $3$ conditions place restrictions on the generating distribution and prior, while the following three do the same for the sampling design distribution. Suppose we have a  sequence, $\xi_{N_{\nu}} \downarrow 0$ and $N_{\nu}\xi^{2}_{N_{\nu}}\uparrow\infty$  and $n_{\nu}\xi^{2}_{N_{\nu}}\uparrow\infty$ as $\nu\in\mathbb{Z}^{+}~\uparrow\infty$ and any constant, $C >0$,

\begin{description}
\item[(A1)\label{existtests}] (Local entropy condition - Size of model)
        \begin{equation*}
        \mathop{\sup}_{\xi > \xi_{N_{\nu}}}\log N\left(\xi/36,\{\lambda\in\Lambda_{N_{\nu}}: d_{N_{\nu}}\left(P_{\lambda},P_{\lambda_{0}}\right) < \xi\},d_{N_{\nu}}\right) \leq N_{\nu} \xi_{N_{\nu}}^{2},
        \end{equation*}
\item[(A2)\label{sizespace}] (Size of space)
        \begin{equation*}
        \displaystyle\Pi\left(\Lambda\backslash\Lambda_{N_{\nu}}\right) \leq \exp\left(-N_{\nu}\xi^{2}_{N_{\nu}}\left(2(1+2C)\right)\right)
        \end{equation*}
\item[(A3)\label{priortruth}] (Prior mass covering the truth)
        \begin{equation*}
        \displaystyle\Pi\left(P_{\lambda}:
        \left(-\mathbb{E}_{\lambda_{0}}\log\frac{p_{\lambda}}{p_{\lambda_{0}}}\leq \xi^{2}_{N_{\nu}}\right)\cap \left(\mathbb{E}_{\lambda_{0}}\left[\log\frac{p_{\lambda}}{p_{\lambda_{0}}}\right]^{2}\leq \xi^{2}_{N_{\nu}} \right)\right) \geq \exp\left(-N_{\nu}\xi^{2}_{N_{\nu}}C\right)
        \end{equation*}
\item[(A4)\label{bounded}] (Non-zero Inclusion Probabilities)
        \begin{equation*}
        \displaystyle\mathop{\sup}_{\nu}\left[\frac{1}{\displaystyle\mathop{\min}_{i\in U_{\nu}}\pi_{\nu i}^{\lambda}}\right] \leq \gamma, \text{  with $P_{\lambda_{0}}-$probability $1$.}
        \end{equation*}
\item[(A5)\label{independence}] (Asymptotic Independence Condition)
        \begin{equation*}
        \displaystyle\mathop{\limsup}_{\nu\uparrow\infty} \mathop{\max}_{i \neq j\in U_{\nu}}\left\vert\frac{\pi_{\nu ij}}{\pi_{\nu i}\pi_{\nu j}} - 1\right\vert = \order{N_{\nu}^{-1}}, \text{  with $P_{0}-$probability $1$}
        \end{equation*}
        such that $\pi_{\nu ij}$ factors to $\pi_{\nu i}\pi_{\nu j}$ for $N_{\nu}$ sufficiently large where there exists some $C_{3} > 0$,
        \begin{equation*}
        \displaystyle N_{\nu}\mathop{\sup}_{\nu}\mathop{\max}_{i \neq j \in U_{\nu}}\left[\frac{\pi_{\nu ij}}{\pi_{\nu i}\pi_{\nu j}}-1\right] \leq C_{3}.
        \end{equation*}
\item[(A6)\label{fraction}] (Constant Sampling fraction)
        For some constant, $f \in(0,1)$, that we term the ``sampling fraction",
        \begin{equation*}
        \mathop{\limsup}_{\nu}\displaystyle\biggl\vert\frac{n_{\nu}}{N_{\nu}} - f\biggl\vert = \order{1}, \text{  with $P_{0}-$probability $1$.}
        \end{equation*}
\item[(A7)\label{pointconverge}] (Convergence of the Point Estimate)
        \begin{equation*}
         \displaystyle\mathop{\limsup}_{\nu\uparrow\infty} \mathop{\max}_{i \in U_{\nu}}\left\vert \mathbb{E}_{\lambda}\left[\pi_{\nu i}\right] - \mathbb{E}_{\lambda_{0}}\left[\pi_{\nu i}\right] \right\vert = \order{N_{\nu}^{-1}}, \text{  with $P_{0}-$probability $1$.}
        \end{equation*}
\end{description}
Condition~\nameref{existtests} restricts the growth in the size of the model space \citep{Ghosal00convergencerates} by bounding the growth in the logarithm of the covering number,
$N\left(\xi/36,\{\lambda\in\Lambda_{N_{\nu}}: d_{N_{\nu}}\left(P_{\lambda},P_{\lambda_{0}}\right)\right)$,
defined as the \emph{minimum} number of balls of radius $\xi/36$ needed to cover $\left\{P\in\mathcal{P}_{N_{\nu}}: d_{N_{\nu}}\left(P,P_{0}\right) < \xi\right\}$ under distance metric, $d_{N_{\nu}}$.
Condition~\nameref{sizespace} allows, but restricts, the prior mass placed on the uncountable portion of the model space, such that we may direct our inference to an approximating sieve, $\mathcal{P}_{N_{\nu}}$.  This sequence of spaces ``trims" away a portion of the space that is not entropy bounded (in condition~\nameref{existtests}).  Condition~\nameref{priortruth} ensures the prior, $\Pi$, assigns mass to convex balls in the vicinity of $P_{\lambda_{0}}$.

The next three conditions, together, restrict the class of sampling designs under which our result is guaranteed. Condition~\nameref{bounded} requires the sampling design to assign a positive probability for inclusion of every unit in the population because the restriction bounds the sampling inclusion probabilities away from $0$. Since the maximum inclusion probability is $1$, the bound, $\gamma \geq 1$.  Unlike in \citet{savitsky2016bayesian}, however, $\pi_{\nu i}^{\lambda}$ is a smoothed model estimator under our fully Bayes construction, which discards the variation in $\pi_{\nu i}$ that is unrelated to $y_{\nu i}$, such that the value of $\gamma$, in practice, is expected to be lower than in \citet{savitsky2016bayesian}. Condition~\nameref{independence} restricts the result to sampling designs where the dependence among lowest-level sampled units attenuates to $0$ as $\nu\uparrow\infty$. Dependence in a multistage design is driven the higher level sampling stages; for example, PSUs.  Since the number of PSUs increases in the limit of $N_{\nu}$, this condition is not very restrictive and admits nearly all sampling designs used, in practice. Condition~\nameref{fraction} ensures that the observed sample size, $n_{\nu}$, limits to $\infty$ along with the size of the partially-observed finite population, $N_{\nu}$. The denominator of our Bayes rule posterior estimator of Equation~\eqref{bayesrule} is a conditional expectation with respect to our model, $P_{\lambda}$.  We require convergence of this point estimate in order to achieve the bound specified in condition~\nameref{independence}.  Condition~\nameref{pointconverge} is not needed in \citet{savitsky2016bayesian}, since the inclusion probabilities, $(\pi_{\nu})$, are assumed fixed.

\begin{theorem}
\label{main}
Suppose conditions ~\nameref{existtests}-\nameref{pointconverge} hold.  Then for sets $\Lambda_{N_{\nu}}\subset\Lambda$, constants, $K >0$, and $M$ sufficiently large,
\begin{align}\label{limit}
&\mathbb{E}_{P_{\lambda_{0}},P_{\nu}}\Pi^{\pi}\left(\lambda:d^{\pi}_{N_{\nu}}\left(P_{\lambda},P_{\lambda_{0}}\right) \geq M\xi_{N_{\nu}} \vert x_{\nu 1}\delta_{\nu 1},\ldots,x_{\nu N_{\nu}}\delta_{\nu N_{\nu}}\right) \leq\nonumber\\
&\frac{16\gamma^{2}\left[\gamma+C_{3}\right]}{\left(Kf + 1 - 2\gamma\right)^{2}N_{\nu}\xi_{N_{\nu}}^{2}} + 5\gamma^{2}\exp\left(-\frac{K n_{\nu}\xi_{N_{\nu}}^{2}}{2\gamma}\right),
\end{align}
which tends to $0$ as $\left(n_{\nu}, N_{\nu}\right)\uparrow\infty$.
\end{theorem}

The rate of convergence is injured for a sampling distribution, $P_{\nu}$, that assigns relatively low inclusion
 probabilities to some units in the finite population such that $\gamma$ will be relatively larger.  Our result differs from
\citet{savitsky2016bayesian} in that $5\gamma$ that multiplies the second term is replaced by $5\gamma^2$, here, though the value $\gamma$ is expected to be lower, as earlier discussed, for our modeled estimate, $\pi_{\nu i}^{\lambda}$, than for the raw, $\pi_{\nu i}$. Similarly, the larger the dependence among the finite population unit inclusions induced by $P_{\nu}$, the higher will be $C_{3}$ and the slower will be the rate of contraction.

The proof of our main result proceeds by bounding the numerator of Equation~\eqref{inform_post} (on the set \\
$\left\{\lambda:d^{\pi}_{N_{\nu}}\left(P_{\lambda},P_{\lambda_{0}}\right) \geq M\xi_{N_{\nu}}\right\}$), from above, and the denominator, from below, and is the same as that outlined in \citet{savitsky2016bayesian}. So we do not repeat it here.  The proof relies two enabling results, which express some differences in this work, however, from \citet{savitsky2016bayesian}. So those Lemma are reformulated and proved in Supplementary Material section~\ref{AppEnabling}.

\section{Simulation Study}\label{sec:simulation}
We explore the bias and coverage performance of our fully Bayes formulation and compare it to the plug-in pseudo posterior model under employment of an informative, probability proportional-to-size (PPS) sampling design; first in a simple linear regression and later in a nonlinear, spline model setting. We construct a synthetic population in each iteration of our Monte Carlo simulation, from which we draw two samples:
1. An informative sample (IS) taken under the PPS design; 2. A simple random sample (SRS).
The IS is analyzed using both the fully Bayes and the pseudo posterior methods.  The SRS is estimated using our population model with no needed correction.  The SRS serves as a something of a gold standard comparator to assess the efficacy of bias removal and uncertainty estimations using the methods that correct for informative sampling, which are estimated on the IS.  Of course, an informative sampling design can be more efficient (lower variance) than an SRS, though an IS is often constructed as much for convenience (and cost) as for efficiency.  We estimate bias, mean square error (MSE), coverage of central 95\% credible interval (CI) and its average length by repeating this this process $M=1,000$ times in a Monte Carlo framework. The three analysis methods (that we label, ``Full" (Fully Bayesian), ``Pseudo" (Pseudo Posterior) and ``SRS" (simple random sampling)) will use the same population distribution for $y_i\mid \bth,\dots $ and prior for $\bth$.
The true value of the generic simulation model parameter, $\eta$, is denoted by $\eta^{\mbox{\tiny{TRUE}}}$.

\subsection{Simple Linear Regression (SLR): PPS Design}
We focus on estimating the population generating for the slope coefficient, $\beta_1$, specified for the linear regression framework in Section~\ref{subsec:LRM}.
The fully Bayes estimation model assumes
a lognormal conditional likelihood for $\pi_i\mid y_i,\cdots$, which is right skewed.
We explore three simulation scenarios;
In the first two, the true generating distribution of the inclusion probabilities
is a gamma distribution and, therefore, skewed as in the the fully Bayes analysis model.
In the third simulation scenario, we explore the robustness of the fully Bayes model to this assumption
by generating the inclusion probabilities from a (symmetric) normal distribution.
The difference between the first and second scenarios is that we vary the rate hyperparameter of the gamma
distribution to induce more variance (and skewness) into the generated $(\pi_{i})$ in the second scenario.

\subsubsection{Scenarios SLR: $\pi$-skewed.}\label{subsubsec:SLRpi-skewed}
We call the first two simulation scenarios
``SLR: $\pi$-skewed with low variance" and
``SLR: $\pi$-skewed with high variance".
Both set $\pi_i\iid \hbox{gamma}(2,rate=b_\pi^{\mbox{\tiny{TRUE}}})$ (See 1 (b) below) where
$b_\pi^{\mbox{\tiny{TRUE}}}=2$, in the former, and $b_\pi^{\mbox{\tiny{TRUE}}}=1$, in the latter.
We set $M=10^3$ to be the total number of synthetic population datasets or Monte Carlo iterations, and $N=10^5$ to
be the number of individuals / units in the population, and we set $n=10^3$ to be the sample size.

\noindent Let $\bth^{\mbox{\tiny{TRUE}}}=(\beta_0^{\mbox{\tiny{TRUE}}},\beta^{\mbox{\tiny{TRUE}}}_1,\beta^{\mbox{\tiny{TRUE}}}_2,\sigma_y^{2,{\mbox{\tiny{TRUE}}}})=(0,1,1,0.1^2)$ under the following simulation procedure:
\begin{itemize}
\item  For $m=1,\dots,M$, Monte Carlo iterations
\begin{enumerate}
      \item Generate the population, \ie, for $i=1,\dots,N$,
 	   \begin{enumerate}
      \item    Draw $\bxy_{i} \iid\hbox{uniform}{(0,1)}$
      \item Draw $\pi_i\iid \text{gamma}(2,rate=b_\pi^{\mbox{\tiny{TRUE}}}) $ (which produces unnormalized $\pi_{i}$).
      \item	Generate the population response:
       $$y_i\mid \bxy_{i},{\pi_{i}},\bth^{\mbox{\tiny{TRUE}}}\sim
        \hbox{normal}( \beta_1^{\mbox{\tiny{TRUE}}} \bxy_{i}+\beta_2^{\mbox{\tiny{TRUE}}}{\pi_{i}},\sigma_y^{2,{\mbox{\tiny{TRUE}}}})$$
       \end{enumerate}
     \item Draw two samples of size  $n=10^3$:
     \begin{enumerate}
     \item Take an IS with,
     ${\mbox{Pr}}[(y_i,\pi_i)\in \hbox{sample}]=\pi_i/\sum_{i^\prime=1}^N \pi_{i^\prime}$.
     \item Take an SRS sample.
     \end{enumerate}
     \item Conduct estimation using the priors given in Equation~\eqref{eq:priors}.
     \begin{enumerate}
     \item The informative sample is estimated using,
     \begin{enumerate}
     \item the fully Bayes linear regression approach outlined in Subsection \ref{subsec:LRM} with parameters,
     $\bth=(\beta_0,\beta_1,\sigma^2_y)$, $\bka=(\kappa_y,\sigma_\pi^2)$, where we note there is no $\bxp$ or $\bka_x$.
      \item the pseudo posterior linear regression formulation.
      \end{enumerate}
       \item The SRS is estimated using the linear regression population model of Equation \eqref{eq:SLR_likelihood}.
     \end{enumerate}
\item
\begin{enumerate}
\item
        Store the posterior expected values of $\beta_1$ under the analyses 3(a) i, 3(a) ii and 3(b).

\item  Also compute the central $95\%$ credible interval (CI) for $\beta_1$ under the three analyses and store their lengths and indicator of whether they contain $\beta_1^{\mbox{\tiny{TRUE}}}$.
 \end{enumerate}
\end{enumerate}
\item Using the values stored in step 4, (a) estimate the Bias and MSE of the estimate of $\beta_1^{\mbox{\tiny{TRUE}}}$ and, (b), estimate the coverage and average length of the central 95\% credible interval (CI) for $\beta_1^{\mbox{\tiny{TRUE}}}$.
\end{itemize}

Note that the synthetic data generating likelihood, $\pi_i\mid y_i,\bxp_i,\bka$,  is \emph{not} the lognormal distribution for the population estimation model; that is, the fully Bayes (estimation) model is misspecified.
Also note that the fully Bayes population  model in (6)
is misspecified under the three analyses. More specifically, the expected value
$y_i\mid \bxy_i,\bth$, after integrating out $\pi_i$, matches that corresponding to a
simple linear regression model with intercept $\beta_0^{\mbox{\tiny{TRUE}}}+E(\pi_i)=0+2/b_\pi^{\mbox{\tiny{TRUE}}}$, slope coefficient
$\beta_1^{\mbox{\tiny{TRUE}}}=1$, but the error, $y_i-E(y_i\mid \bxy_i,\bth)$, though having mean zero and constant variance, $var(\pi_i)+\sigma_y^{2,{\mbox{\tiny{TRUE}}}}=2/(b_\pi^{\mbox{\tiny{TRUE}}})^2+0.1^2$, is \emph{not} normally distributed.

\begin{figure}
\begin{center}
\includegraphics[width=70mm,angle=0]{\plotpath/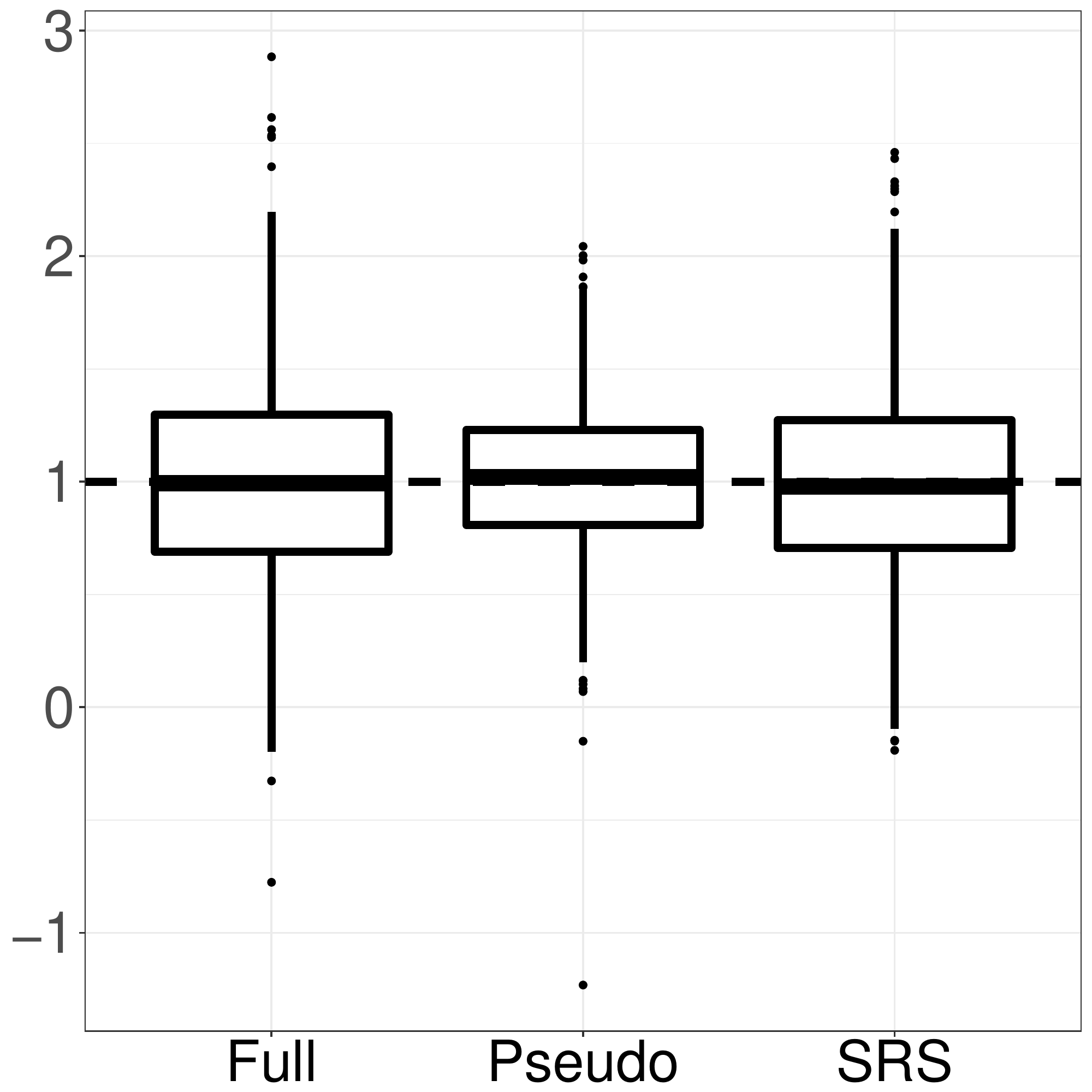}\\%
\end{center}
\vskip-5mm
\caption{\label{fig:SLRpi-skewed} Estimation of $\beta_1^{\mbox{\tiny{TRUE}}}$ under simulation scenario, SLR: $\pi$ skewed with low variance.
Here, $\pi_i\iid \hbox{gamma}(2,b_\pi^{\mbox{\tiny{TRUE}}}=2)$ in step 1(a) in Section~\ref{subsubsec:SLRpi-skewed}.
Box plots of posterior expected values of $\beta_1$ are displayed under IS for the fully Bayes and
pseudo posterior approaches and under SRS. The horizontal dashed line represents simulation value $\beta_1^{\mbox{\tiny{TRUE}}}=1$.}
\end{figure}
\FloatBarrier

Figure \ref{fig:SLRpi-skewed} displays boxplots of the $M$ posterior expected values of $\beta_1$ under each analysis
under scenario SLR: $\pi$-skewed with low variance. Table \ref{table:SLRpi-skewed} presents the bias, MSE, coverage of central 95\%
CI and its average length under both scenarios. The figure and the left-hand columns 2 to 4 of the table show that when the true variance of $\pi$ is relatively \emph{low} the pseudo posterior yields a good point estimator of $\beta_1$ with MSE lower even than under SRS.
There is a trade-off of some bias for improved efficiency. The right-hand columns of the table, reporting the results under scenario SLR: $\pi$-skewed with relatively \emph{high} variance, show that this property does not hold when the variance of $\pi$ is large;
the MSE for the pseudo posterior  is now the highest, as is the bias, so there is no trade-off of one for the other.
 In other words,
the point estimate of the fully Bayes model is robust against high variability of the inclusion probabilities while pseudo posterior estimator is not.
In both scenarios,
the average length of the CI is notably shorter under SRS than for the other two approaches under an IS because there is more variation in the realized samples drawn under an informative design, $P_{\nu}$, than under SRS.
The pseudo posterior underestimates the uncertainty (of the point estimate for $\beta_1$); its CI fails to capture the nominal $95\%$ coverage.  In contrast, the fully Bayes model estimates the uncertainty appropriately; its CI maintains a coverage similar to the model under SRS.
\begin{table}[ht]
\centering
\begin{tabular}{r| rrr |rrr}
True variance & \multicolumn{3}{c}{Low ($b^{\mbox{\tiny{TRUE}}}_\pi=2$)} 									&\multicolumn{3}{c}{High ($b_\pi^{\mbox{\tiny{TRUE}}}=1$)}\\
  of $\pi_i$ & Full & Pseudo & SRS 												& Full & Pseudo & SRS \\
 \hline
  Bias & -0.003 & 0.013 & -0.002 										& -0.007 & {1.031} & 0.005\\
  MSE & 0.219 & 0.109 & 0.183 										& 0.509 & {1.278} & 0.374\\
  95\% CI coverage& 0.944 &{0.861} & 0.947  		& 0.926 & 0.839 & 0.933\\
   95\% CI length& 1.770 & 1.663 & {0.952} 		& 2.472 & 2.227 & 2.212\\
\end{tabular}
\caption{\label{table:SLRpi-skewed} Bias, MSE, coverage of central 95\% CI and
its average length with three analyses under
simulation scenarios SLR: $\pi$-skewed with low variance and with high variance.
}
\end{table}

\subsubsection{Scenario SLR: $\pi$-symmetric.}
We explore the robustness of our approach in the case where the inclusion
probabilities are generated as symmetric but modeled as skewed.
To do so, we repeat the simulation study in subsection
\ref{subsubsec:SLRpi-skewed}, but now with $\pi_i\iid\text{normal}^{+}(1,\sigma_\pi^{2,{\mbox{\tiny{TRUE}}}}=0.1^2)$ in 1 (b).
We call this simulation scenario: ``SLR: $\pi$-symmetric".
Notice that the fully Bayes approach, as in simulation scenario scenario SLR: $\pi$-skewed,
misspecifies the distribution of $\pi_i\mid y_i,\bka$. In contrast to scenario SLR: $\pi$-skewed,
the regression model for $y_i\mid \bxy_i,\bth$ is correctly specified; in particular the distribution of the error is normally distributed.

\begin{figure}
\begin{center}
\includegraphics [width=70mm,angle=0]{\plotpath{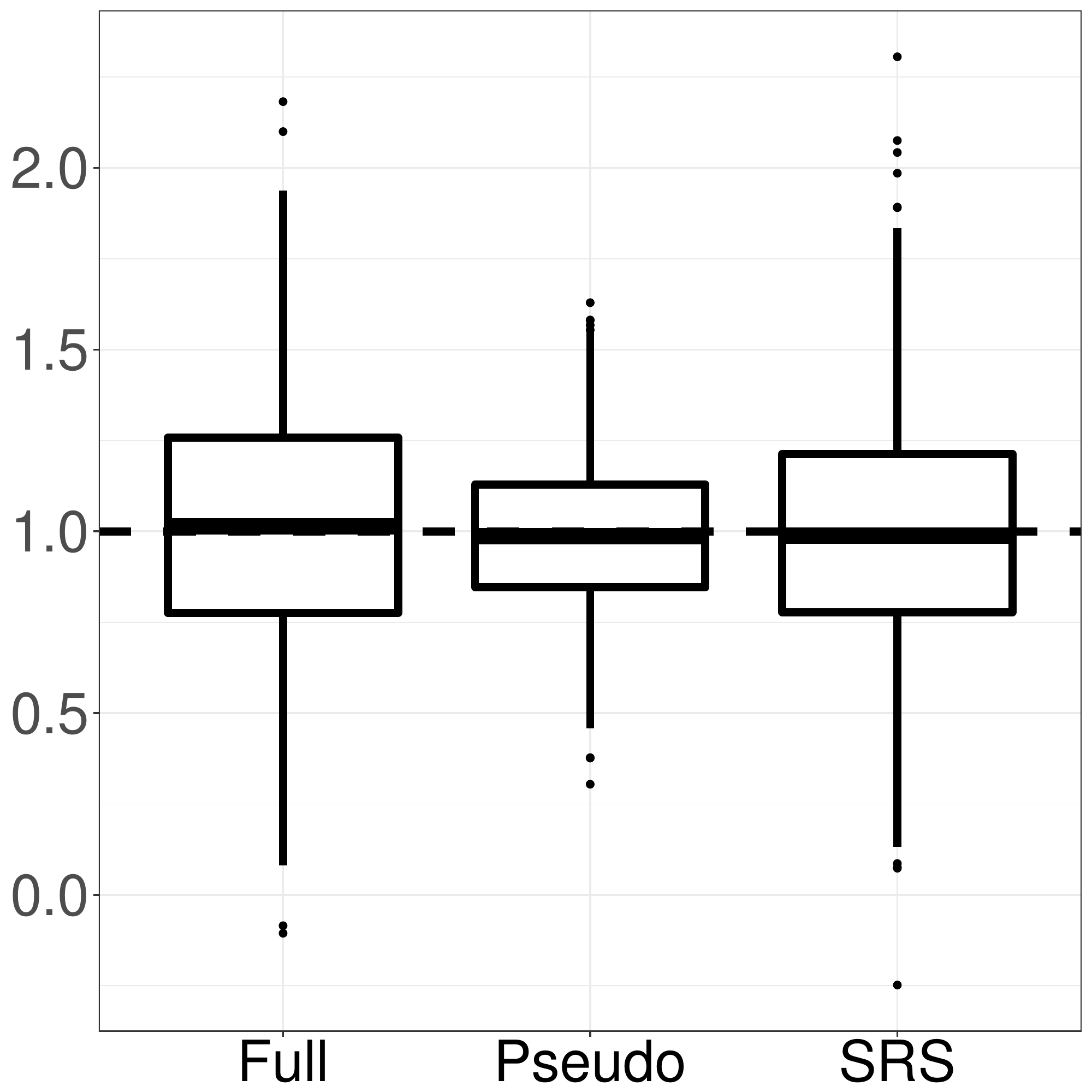}}\\
\end{center}
\vskip-5mm
\caption{\label{fig:SLRpi-symmetric} Estimation of $\beta_1$ under simulation scenario, SLR $\pi$-symmetric. Box plots of posterior expected values of $\beta_1$ are displayed under IS for the fully Bayes and
pseudo posterior approaches and under SRS. The horizontal, dashed line represents the simulation value of $\beta_1^{\mbox{\tiny{TRUE}}}=1$.}
\end{figure}

Figure~\ref{fig:SLRpi-symmetric} and Table~\ref{table:SLRpi-symmetric}
are constructed in the same formats as Figure \ref{fig:SLRpi-skewed} and Table \ref{table:SLRpi-skewed},
for scenario: SLR: $\pi$-symmetric.  The three methods perform similarly in terms of bias, MSE and coverage, even though, by design, the population model for the inclusion probabilities is misspecified under the fully Bayes method because we have drawn the inclusion probabilities from a symmetric distribution and modeled them with a skewed distribution.
As in the previous scenario, SLR:$\pi$-skewed, the average length of the CI is
notably shorter under SRS than under the other two approaches. That is, both, the fully Bayes and the
pseudo posterior approaches maintain nominal coverage at the cost of wider credible intervals.

\FloatBarrier

\begin{table}[ht]
\centering
\begin{tabular}{r|rrr }
 & Full & Pseudo & SRS \\											
  \hline
 Bias & 0.013 & -0.011 & 0.002 				\\
 MSE & 0.123 & 0.042 & 0.115                      \\
 95\% CI coverage & 0.957 & 0.948 & 0.950\\
  95\% CI length & 1.373 & 1.370 & 0.795\\

  \end{tabular}

\caption{\label{table:SLRpi-symmetric} Bias, MSE, coverage of central 95\% CI and its average length
with three analyses under simulation scenario SLR: $\pi$-symmetric.
}
\end{table}
\FloatBarrier

\subsubsection{Scenario non-linear: $\pi$-skewed.}\label{subsubsec:nonpi-skewed}
We conclude our simulation studies by exploring the situation where the relationship between the predictor and the response is \emph{not} linear.  As before, the superindex ${\mbox{\tiny{TRUE}}}$ indicates the simulation model parameters.
To ease the computational burden, we simulate only one synthetic population dataset with $N=10^5$
individuals. 
We run this simulation twice, first with sample size $n=100$ and last with $n=1000$.
The objective is to estimate the curve
$E(y\mid \bxy)$ on a regular grid, $\bxy\in(0,2)$. The analysis model is the splines regression model described in Subsection \ref{subsection:spline}.
\begin{itemize}
\item Generate the population (once)
 	\begin{enumerate}
      \item    Draw $\bxy_{i} \iid\hbox{uniform}{(0,2)}$
      \item Draw $\pi_i\iid \text{gamma}(2,rate=1) $
      \item	Generate the population response:
       $$
       \begin{array}{ll}
       y_i\mid \bxy_{i},\pi_{i},(\beta^{\mbox{\tiny{TRUE}}}_{\bxy},\beta^{\mbox{\tiny{TRUE}}}_\pi,\beta^{\mbox{\tiny{TRUE}}}_{\bxy,\bxy},\sigma_y^{2,{\mbox{\tiny{TRUE}}}})\sim &
 \hbox{normal} (\beta_{\bxy}^{\mbox{\tiny{TRUE}}} \bxy_i+\\
 &\beta_\pi^{\mbox{\tiny{TRUE}}} \pi_{i}+
 \beta_{\bxy,\bxy}^{\mbox{\tiny{TRUE}}} \bxy_i^2,\\&\sigma_y^{2,{\mbox{\tiny{TRUE}}}}),
 \end{array}
 $$
 with $(\beta^{\mbox{\tiny{TRUE}}}_{\bxy},\beta^{\mbox{\tiny{TRUE}}}_\pi,\beta^{\mbox{\tiny{TRUE}}}_{\bxy,\bxy},\sigma_y^{2,{\mbox{\tiny{TRUE}}}})=(1,1,-0.5,0.1^2)$
 \end{enumerate}

    \item For $m=1,\dots,M$:
   \begin{enumerate}
       \item       Generate an IS and SRS of size $n$ in the same fashion as in the SLR simulation study and, subsequently, estimate the model described in Section~\ref{subsection:spline},
      using cubic B-splines, 
     with the first and last internal knots equal to, respectively,
     the minimum and the maximum of the $\bxy_i$s in the sample.
\item  Compute the central $95\%$ credible intervals for $E(y\mid \bxy)=B(\bxy)E(\bbe\mid \by,\bpi)$, pointwise on a grid, $i.e.$, $\bxy=0,1/40,2/40,\dots,$ $80/40$,  under the three analyses and store their credible interval lengths and an indicator for whether or not
they contain the true curve given by
$$\beta_{\bxy}^{\mbox{\tiny{TRUE}}} \bxy+\beta_\pi^{\mbox{\tiny{TRUE}}} E(\pi_{i})+\beta_{\bxy,\bxy}^{\mbox{\tiny{TRUE}}} \bxy^2=
\bxy+2-0.5\bxy^2
  $$
\end{enumerate}
\item Based on the values stored in step 3 (a), estimate the (pointwise) Bias and MSE curves
and based
        on the values stored in step 3 (b), estimate the coverage and average length of the central 95\% CI
        for the curve $E(y\mid \bxy)$ with $\bxy\in(0,2)$.
\end{itemize}

\begin{figure}
\begin{center}
\includegraphics[width=100mm,angle=0]{\plotpath/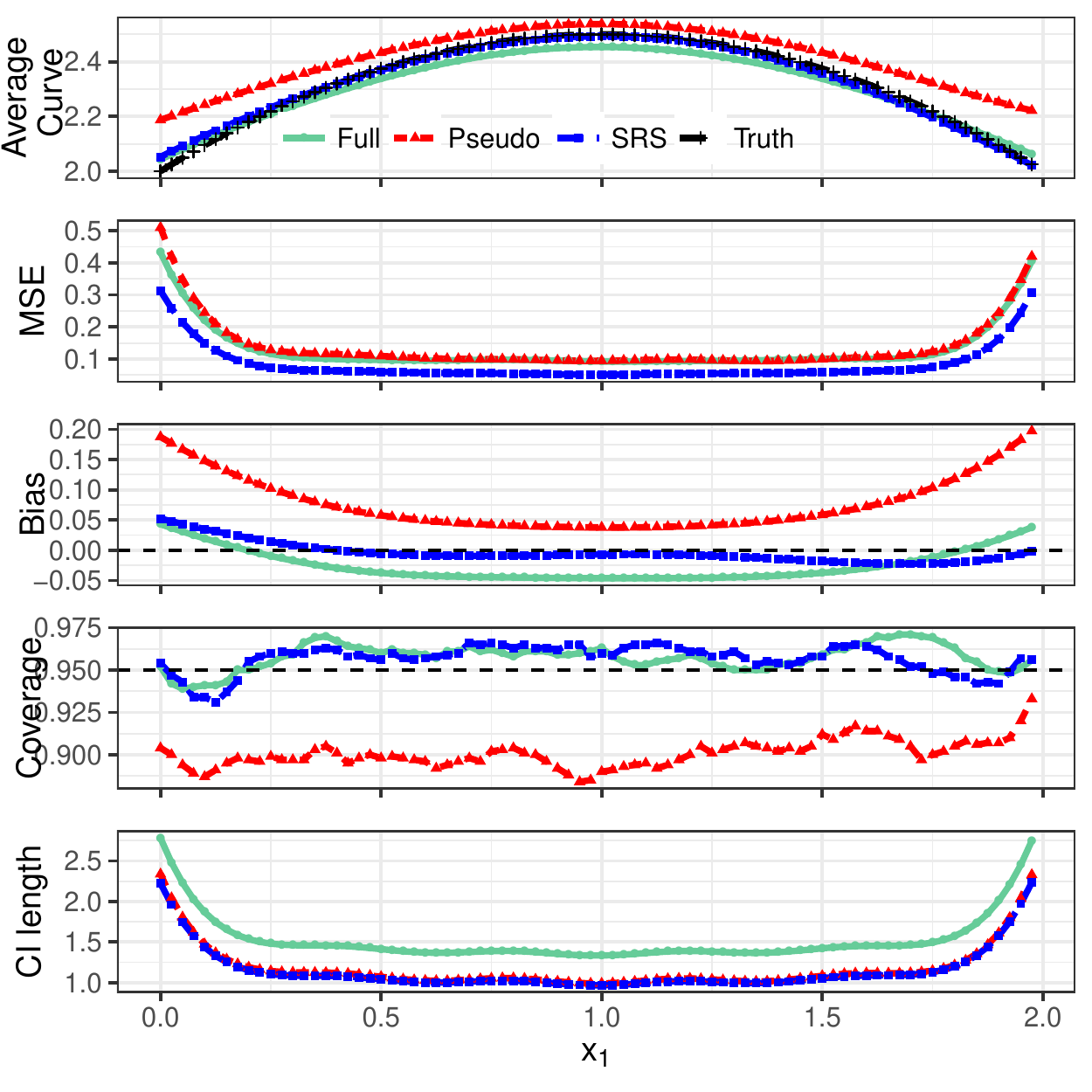} \\
\end{center}
\vskip-5mm
\caption{\label{fig:MSEBiasx2_gamma_21_n_100}
Simulation scenario non-linear: $\pi$-skewed with sample size $n=100$. From top to bottom,
average estimated curve, MSE, bias, coverage probability of 95\% central credible intervals and their average length.
The black curve in the upper panel is the true curve.  For visual purposes the horizontal (discontinuous)
lines at 0 and 0.95 in third and fourth panels, respectively, are depicted.
}
\end{figure}

\begin{figure}
\begin{center}
\includegraphics[width=100mm,angle=0]{\plotpath/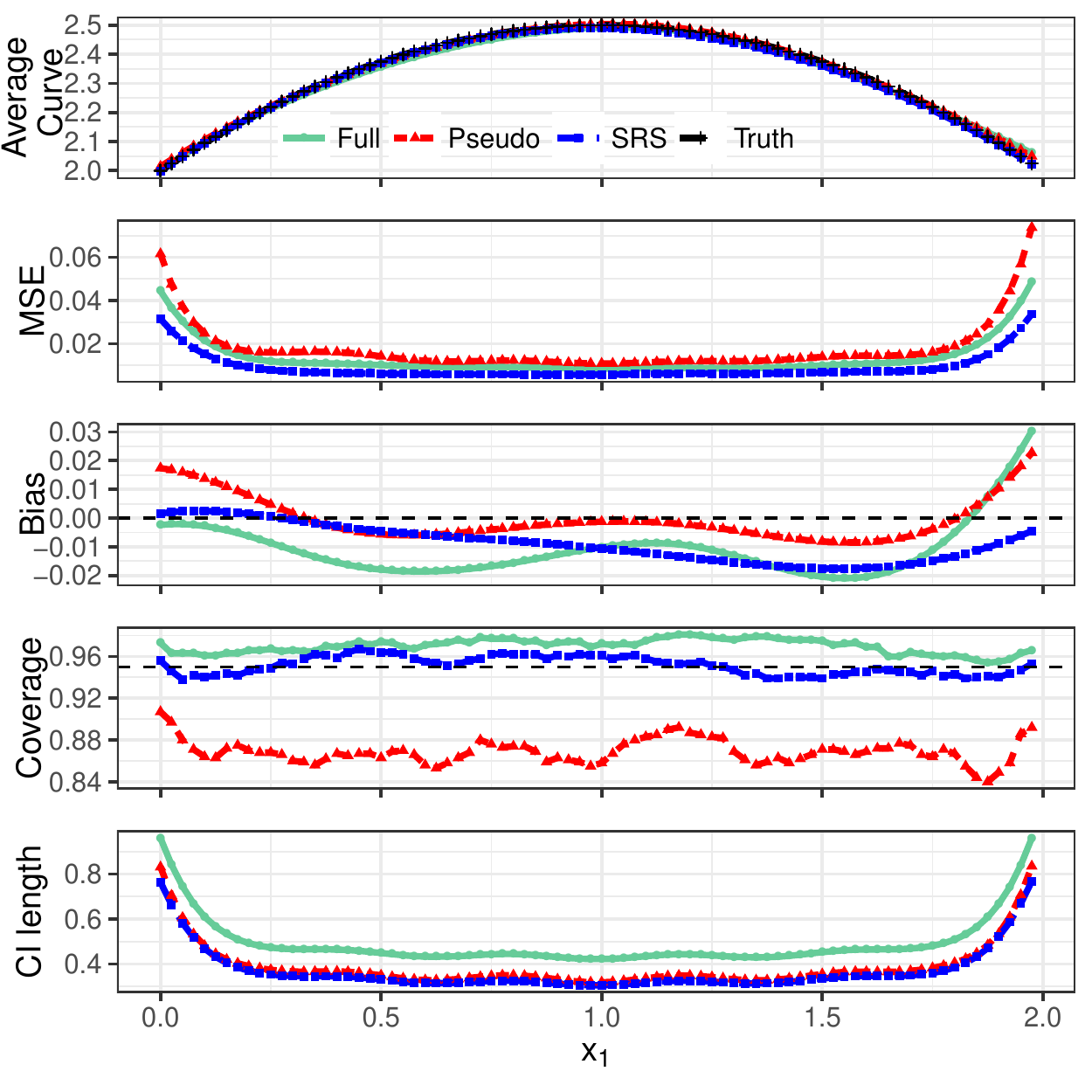} \\
\end{center}
\vskip-5mm
\caption{
\label{fig:MSEBiasx2gamma21_n_1000}
Simulation scenario non-linear:$\pi$-skewed with sample size $n=1000$.
Same as Figure \ref{fig:MSEBiasx2_gamma_21_n_100} but with $n=1000$.
}
\end{figure}

Figure \ref{fig:MSEBiasx2_gamma_21_n_100} shows the results of the simulation under this scenario with sample size,
$n=100$.
The first plot panel (from the top) displays the simulation curve and those
estimated under the three methods. The next four panels (from top-to-bottom) display the MSE, Bias, coverage of central 95\% CI and the average length of the CI for the three methods, respectively. The SRS method performs uniformly the best in terms of fit, bias, MSE and coverage. It is followed by the fully Bayes approach. As in scenario SLR:$\pi$-skewed the pseudo posterior approach fails to maintain a 95\% coverage for
its central 95\% CI. The bottom panel shows that the cost of having an IS is a wider CI for the same coverage due to the larger variation in information about the population in samples drawn under an informative design, as earlier discussed.
Figure
\ref{fig:MSEBiasx2gamma21_n_1000} is the same as Figure
\ref{fig:MSEBiasx2_gamma_21_n_100} after increasing the sample size to $n=1000$.
The results
are qualitatively the same as when $n=100$
except that the pseudo posterior performs better in terms of bias; in agreement with
the result that the pseudo posterior approach produces asymptotically
unbiased estimators \citep{savitsky2016bayesian}.

\FloatBarrier



\section{Application}\label{sec:application}
As an illustration we explore the relationship between caffeine and systolic blood pressure (SBP),
applying our methodology to data from the 2013-2014 National Health and Nutrition Examination Survey (NHANES).
We will perform two analyses. In the first analysis we do not adjust for any other covariate and in the second we adjust for age and gender.

Due in part to technical, budget and logistic considerations, the survey design of the NHANES is complex.   NHANES is designed to assess the health and nutritional status of adults and children in the United States \citep{CDCA}. Data are collected from a sample of individuals from the non-institutionalized, civilian US population. Although nationally representative, the NHANES is designed to oversample specific subpopulations of interest (e.g. minorities, low income groups, pregnant women) for population-based studies using a complex, multistage, cluster sampling design \citep{CDCB}. To correct for informativeness (partly induced by the correlation between response values and group memberships), the NHANES survey data are released with observation-indexed sampling weights (based on marginal inclusion probabilities) corresponding to the sampled participants. These sampling weights are computed based on the sampling design (i.e. probability of the individual of being included in the sample), non-response and the masking of the individual for confidentiality \citep{CDCD}.

NHANES participants are required to visit a mobile examination center where Health and nutrition information are collected during these visits.
In particular, SBP is measured for participants of ages $8$ years and older. A 24-hour dietary recall interview is taken. The results of this interview are used to estimate the participant's intakes of nutrients during the 24-hour period preceding their interview (from the period of midnight to midnight). Our dataset consists of systolic blood pressure (SBP) (mm Hg) measurements of $n=6847$ participants 8 or older and 24-hour caffeine consumption (mg), along with their sampling weights, as estimated from the first 24-hour dietary recall interview. See Supplementary Material
\ref{sub:datapreprocess} for details regarding our data preprocessing.

We fit the spline basis model  introduced in  Subsection \ref{subsection:spline} with response, $y=\log(\hbox{SBP})$,
inclusion probability, $\pi$, designed to be proportional to the inverse of the NHANES sampling weight, and
predictor, $x=\log(\hbox{caffeine consumption}+1) $.
We fit  both the fully Bayesian (Full) and the pseudo posterior (Pseudo) models.  Additionally,
against the recommendation stated in NHANES, we fit the splines model for the population that ignores IS (equivalent to the pseudo posterior model with all sampling weights equal to 1).  The cubic B-spline model is constructed with the priors and hyperparameters described in Section \ref{subsection:spline}.

\begin{figure}
\begin{center}
\includegraphics [width=80mm,angle=0]{\plotpath{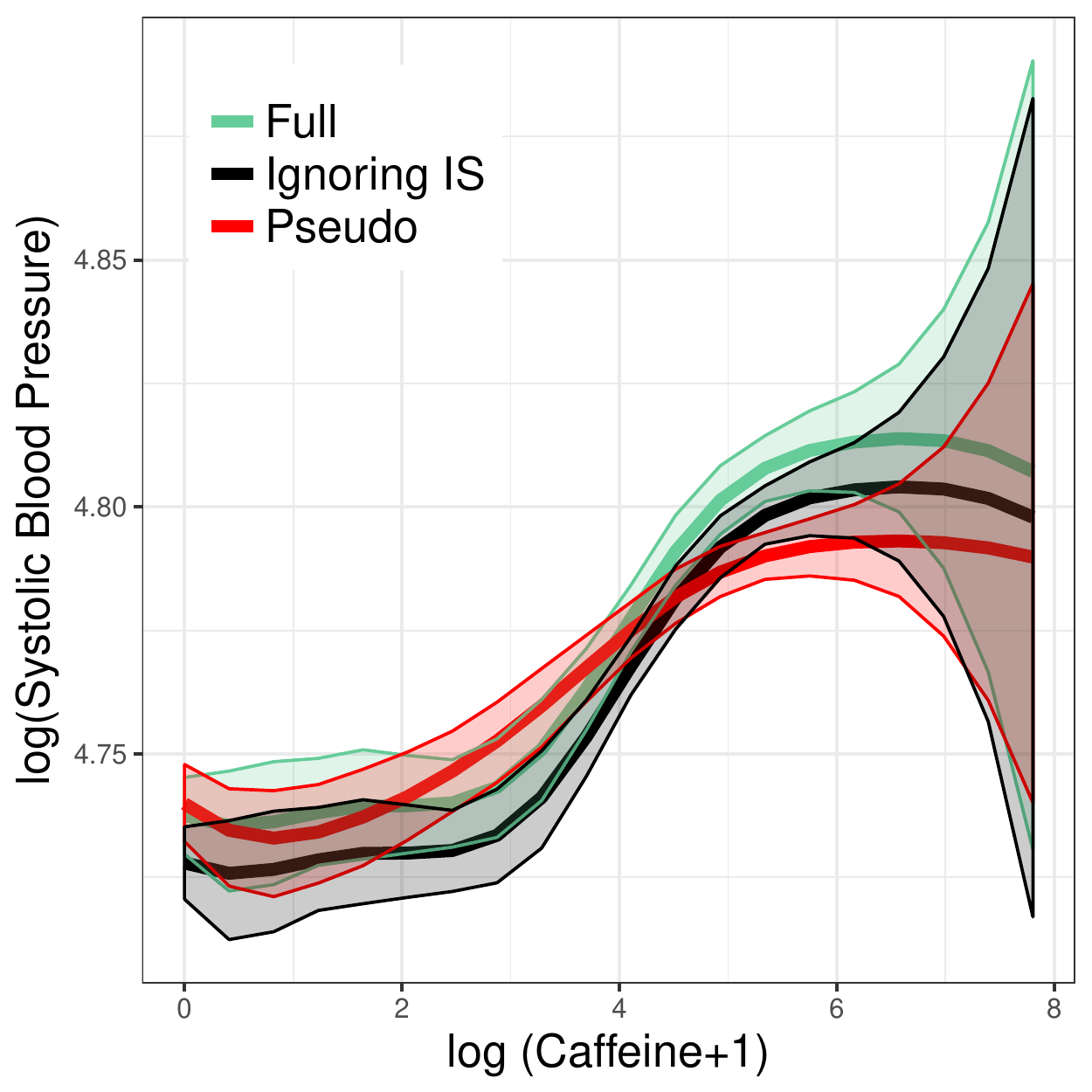}}\\
\end{center}
\vskip-5mm
\caption{\label{fig:logcaffeinevsSysBP} Caffeine consumption (mg) \textit{v.s.} SBP (mm Hg) under the Full, Pseudo and Ignoring IS models.}
\end{figure}

After a burn-in period of $10^4$ iterations, a Monte Carlo posterior sample of size $10^4$  was retained for each of the model parameters and used to estimate $E(y\mid x)$, together with 95\% credible intervals on a pointwise basis under the three models. Figure \ref{fig:logcaffeinevsSysBP} displays these curves. There is a positive relationship between caffeine consumption and blood pressure that levels off in higher ranges of caffeine consumption.  The Pseudo posterior estimates a gradually increasing relationship, while the Fully Bayes estimates $3$ distinct regions in the support of log(caffeine$+1$), where each expresses a different sensitivity to log(SBP); in particular, the credible intervals for these two models do not overlap in parts of the second region (from left-to-right) in a range of $2.5 \leq x \leq 4$, where the Fully Bayes model shows a notably greater sensitivity.  In this case, then, the inferences performed under these two models would differ.

The resulting negative sign of $\kappa_y$, with central 95\% credible interval (-0.64,-0.35)
and $\mbox{Pr}(\kappa_y<0\mid \hbox{data})\approx 1$, in the model for $\pi_i\mid y_i,x_i$ indicates that the sample design is informative for the logarithm of SBP (when not conditioned on predictors). The higher the value of the response the lower is the inclusion probability.

\begin{figure}
\begin{center}
\includegraphics [width=60mm,angle=0]{\plotpath{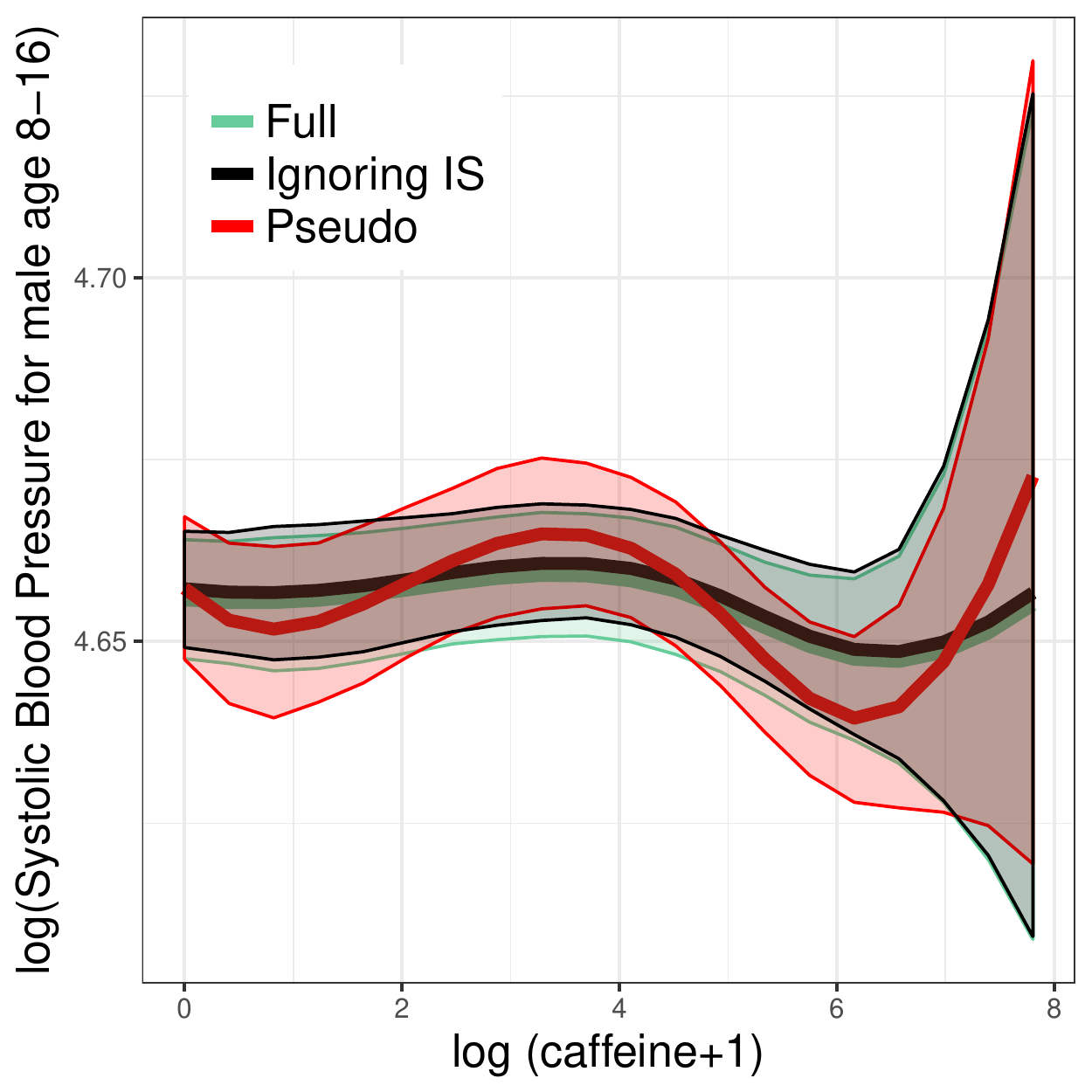}}\\
\end{center}
\caption{\label{fig:caffeinevsSBPcontrollingforAgeandGender}
Caffeine consumption (mg) \textit{v.s.} SBP (mm Hg), or males 8-16 years old, under the Full, Pseudo and Ignoring IS models.
}
\end{figure}

We next proceed to check if this positive association between SBP and caffeine consumption still holds after controlling for
age and gender.  We categorized the age into seven groups
 as shown in Table
\ref{table:caffeinewithageandgender}. The reference categories were the 8-17 age group and the male gender.
We adjust the cubic B-spline model
$$y_i\sim \hbox{normal}\left(\mu(\bxy_i)+\beta_{age(i)}^{age}+\beta_{gender(i)}^{gender},\sigma_y^2\right)$$
with $\mu(\bxy_i)$ defined in
Equation~\eqref{eq:Splinemodel}. $\beta_{age(i)}^{age}$ and $\beta_{gender(i)}^{gender}$
are the regression coefficients associated with the age group and gender of the participant $i$ for groupings,
$age(i)\in \{1,\dots,7\}$ and $gender(i)\in \{1,2\}$.
The regression coefficients associated to the reference groups are set to $0$
($\beta^{age}_1\equiv \beta^{gender}_1 \equiv 0$).
We employ vague priors
$\beta^{age}_2,\dots,\beta^{age}_7,\beta^{gender}_2\iid \hbox{normal}(0,10^4)$.
Age group and gender were also included as covariates in $\pi_i\mid y_i,\bxp_i$, where $\bxp_i$ is a vector of dimension $9$, by incorporating for an intercept, $x\equiv\log(\hbox{caffeine}+1)$, seven age groups and gender.
The estimated curves along with their 95\% credible intervals for the reference group (males between 8 and 16 years old) under the three models
are shown in Figure~\ref{fig:caffeinevsSBPcontrollingforAgeandGender}.
The positive association between caffeine consumption and SBP vanishes when controlling for age and gender.
The estimated curves under the Fully Bayes and Pseudo posterior models differ in shape but their bands overlap across all the range of $x$. The posterior mean curve for the fully Bayes model is, however, more smoothly centered on the horizontal line to more strongly indicate little-to-no association between systolic blood pressure and caffeine consumption for the reference group.  The credible intervals for the Fully Bayes model and that ignoring IS are almost perfectly overlapping, which suggests a non-informative sampling design. The central 95\% credible interval for $\kappa_y$, $(-.06,0.28)$, contains zero, confirming that the sampling design is non-informative for SBP when controlling for age and gender.
Table~\ref{table:caffeinewithageandgender}
depicts the  posterior mean, standard deviation and 2.5\% and 97.5\% quantiles
of the marginal posterior distribution of the Fully Bayes, adjusting for age and gender, model parameters.
The regression coefficients in the conditional model for $y_i\mid\bxy_i$ indicate that
mean SBP for females is lower and SBP increases with age.

 When the sampling design is not informative for the chosen response (given the available predictors), the fully Bayes model provides similar inference than under SRS. By contrast, the pseudo posterior estimator is notably noisier than under SRS, which in some cases may lead to incorrect inference.

\cite{steffen2012effect} performed a meta-analysis of randomized clinical trials and prospective studies and concluded that coffee consumption was not associated with a significant change in SBP. The Fully Bayes and ``Ignoring IS' curves in Figure~\ref{fig:caffeinevsSBPcontrollingforAgeandGender} are nearly horizontal, which effectively reproduces this result.  The pseudo posterior result, however, expresses a lot more oscillation around around the horizontal line.

\begin{table}
\begin{center}
\begin{tabular}{rrrrr}
&\textbf{mean} & \textbf{sd} & \textbf{2.5\%} & \textbf{97.5\%} \\
  \hline
  \multicolumn{5}{c}{Parameters of $y_i\mid \cdots$}\\
  \multicolumn{5}{c}{Regression coefficients}\\
Gender (female) &-0.03 & 0.00 & -0.04 & -0.03 \\
  Age 17-24& 0.08 & 0.00 & 0.07 & 0.09 \\ 
  Age 25-34 & 0.09& 0.01 & 0.08 & 0.10 \\ 
  Age 35-39 & 0.12 & 0.01 & 0.11 & 0.13 \\ 
  Age 40-49 & 0.14 & 0.01 & 0.13 & 0.15 \\ 
  Age 50-59 & 0.18 & 0.01 & 0.17 & 0.19 \\ 
  Age 60 or older & 0.24 & 0.00 & 0.23 & 0.25 \\ 
  $\sigma_y$ & 0.11 & 0.00 & 0.11 & 0.12 \\
    \multicolumn{5}{c}{(Caffeine) Spline parameters in $y_i\mid \cdots$}\\
$\beta_1$ & 4.66 & 0.03 & 4.59 & 4.73 \\
$\vdots$\\
  \hline
      \multicolumn{5}{c}{Parameters of $\pi_i\mid y_i,\cdots$}\\
  Intercept &  -0.91 & 0.41 & -1.71 & -0.11 \\
  $\kappa_y$&0.11 & 0.09 & -0.06 & 0.28 \\
  Caffeine & -0.05 & 0.01 & -0.06 & -0.04 \\
  Gender (Female)  & -0.03 & 0.02 & -0.07 & 0.01 \\
 Age 17-24  & -0.55 & 0.04 & -0.62 & -0.48 \\ 
  Age 25-34  & -0.74 & 0.04 & -0.81 & -0.67 \\ 
  Age 35-39  & -0.67 & 0.05 & -0.77 & -0.58 \\ 
  Age 40-49  & -0.67 & 0.04 & -0.75 & -0.59 \\ 
  Age 50-59  & -0.72 & 0.04 & -0.80 & -0.63 \\ 
  Age 60 or older & -0.54 & 0.04 & -0.62 & -0.47 \\
  $\sigma_\pi$ & 0.83 & 0.01 & 0.81 & 0.84 \\
   \hline
\end{tabular}
\end{center}
\caption{\label{table:caffeinewithageandgender}
Fully Bayes model parameter estimates for caffeine consumption versus SBP. (Caffeine) spline parameters $\beta_2,\dots,\beta_8$
and $\sigma_\beta$ are not depicted to save space.}
\end{table}
\FloatBarrier
\section{Discussion}\label{sec:discussion}
We have proposed a fully Bayes approach to incorporate the sampling weights into the estimation of population model
parameters on data collected under an informative sampling design.  The approach does not require extensive knowledge of sampling design, uses only quantities observed in sample, and discards variation in weights not dependent on response.
We have shown via simulation that the method performs as well and often better in terms
of bias, MSE and, the coverage of its central 95\% credible interval than the non-fully Bayesian plug-in pseudo posterior.
Our fully Bayes approach demonstrates a consistently superior ability to accurately measure uncertainty, in contrast
with the pseudo posterior method, which is too confident such that is fails to achieve nominal coverage.  The price to be paid for the achievement of unbiased inference with correct coverage of population model parameters from an informative sample is wider credible intervals to achieve the same coverage relative to simple random sampling.
We applied the Fully Bayes approach to explore the relationship between caffeine consumption and SBP
analyzing NHANES data. We notice a (non linear) positive association between this two variables. This association vanishes when controlling for age and gender. The Fully Bayesian approach estimates a smoother and closer to the horizontal line (indicating no association) caffeine vs SBP curve as compared to the pseudo posterior approach, which better agrees with previous research for the relationship between caffeine consumption and SBP.

The unusual assumption of our approach, from the perspective of the survey sampler, is that it considers the weights (inclusion probabilities) as random. Although the fully Bayes approach requires specification of a conditional distribution for the inclusion probabilities, we have shown that
coverage properties are robust to misspecification of this distribution.
A disadvantage of the proposed method is that it requires a customized posterior sampler because the posterior sampler designed for simple random sampling cannot easily be adapted to IS under the fully Bayes construction.
We cope with this problem by relying on Stan~\citep{carpenter2016stan}.
A related issue is the required integration step for every observation at every iteration of the posterior sampler.
This computation must be performed, numerically, in the general case. Theorem  \ref{th:closeform} provides some conditions
in the sampling distribution and the conditional distribution of the weights given the response
that allow a closed form for the fully Bayes likelihood. We applied the theorem to illustrate a collection of useful population joint distributions over the inclusion probabilities and response that all allow for a closed form for the integration step.
More research is needed to apply our approach to more complicated settings. Future work will focus on performing the computation of this expected value, numerically, or circumventing it by introducing latent random variables.

\section*{Supplementary Material}
Contains:
\begin{enumerate}
\item Proof of Theorem \ref{th:closeform} that specifies conditions under which a closed form is available for the expectation step in  the denominator of the fully Bayes estimator.
\item Derivation for the closed form expression for $\log p_s(\dots)$ in Subsection \ref{subsec:LRM}
\item Enabling Lemmas for Theorem~\ref{main}.
\item Detailed
data preprocessing steps for application in Section \ref{sec:application}.
\end{enumerate}
\par

\begin{frontmatter}
\title{Supplementary Material: Fully Bayesian Estimation Under Informative Sampling}
\runtitle{Supplemental: Fully Bayes Under Informative Sampling}

\begin{aug}
\author{\fnms{Luis G.} \snm{Le\'on-Novelo}\thanksref{t2}\ead[label=e1]{Luis.G.LeonNovelo@uth.tmc.edu}}
\and
\author{\fnms{Terrance D.} \snm{Savitsky}\thanksref{t3}\ead[label=e2]{Savitsky.Terrance@bls.gov}}

\address{1200 Pressler St. Suite E805, Houston, TX, 77030, USA and \\
PSB Suite 1950, 2 Massachusetts Avenue, NE Washington, DC 20212-0001, USA.\\
\printead{e1,e2}}

\thankstext{t2}{Department of Biostatistics and Data Science, 
University of Texas Health Science Center at Houston- School of Public Health}
\thankstext{t3}{Office of Survey Methods Research, Bureau of Labor Statistics}
\runauthor{Le\'on-Novelo and Savitsky}


\end{aug}
\end{frontmatter}

\ref{sec:appendixprooftheoremcloseform}, 
proof of Theorem \ref{th:closeform};
\ref{subsec:suppleLinReglikelihood}, 
closed form expression for $\log p_s(\dots)$ in Subsection \ref{subsec:LRM}; 
\ref{AppEnabling}, enabling Lemmas for Theorem~\ref{main}; 
and,
\ref{sub:datapreprocess}, detailed data preprocessing steps for application in Section \ref{sec:application}.

\section{Proof of Theorem~\ref{th:closeform}}\label{sec:appendixprooftheoremcloseform}
Recall that the vector of covariates $\bx_i$ has been split into two
sets of covariates, $\bxy_i$ and $\bxp_i$, relevant for the distribution of
$y_i\mid \bth,\bxy_i$, and  $\pi_i\mid y_i,\bxp_i,\bka$, respectively. Further recall the definition for
$M_y(\bka;\bxy_i,\bxp_i,\bth):=E_{y\mid \bxy_i,\bth}\left[\exp\left\{\hy(y_i,\bxp_i,\bka)\right\}\right]
$.
Using the fact that  $\upsilon\sim \hbox{lognormal}(m,v^2)\Rightarrow$ $E(\upsilon)=\exp(m+v^2/2)$, we obtain,
\begin{align*}
E_{y\mid \bxy_i,\bth}\left\{ E\left[\pi_i\mid y,\bxp_i,\bka\right]\right\} =&
E_{y\mid\bxy_i,\bth}\left[ \exp\{\hy(y_i,\bxp_i,\bka)+\hmy(\bxp_i,\bka)+\sigma^2_\pi/2        \}\right]\nonumber\\
=&
\exp\left[\hmy(\bxp_i,\bka)+\sigma^2_\pi/2\right]\times E_{y\mid\bxy_i,\bth}\left[ \exp\{\hy(y_i,\bxp_i,\bka)        \}\right]\nonumber\\
=&\exp\left[\hmy(\bxp_i,\bka)+\sigma^2_\pi/2\right]\times M_y(\bka;\bxy_i,\bxp_i,\bth) \nonumber\\
\nonumber
\end{align*}
Noting that
$\hbox{lognormal}(\pi \mid m,v^2)=\hbox{normal}(\log \pi\mid m,v^2)\times 1/\pi$,
plugging the expression above in equation  \eqref{eq:IScorrection}
 we obtain,
\begin{align*}
p_s(y_i,\pi_i\mid \bxy_i,\bxp_i,\bth,\bka)
=&
\frac{
	\pi_i p(\pi_i\mid y_i,\bxp_i,\bka)
}{
E_{y\mid \bxy_i,\bth}\left\{ E\left[\pi^\prime_i\mid y,\bxp_i,\bka\right]\right\}
}
 p(y_i\mid \bxy_i,\bth) \nonumber\\
 =&\frac{\pi_i\times \hbox{lognormal}\left(\pi_i\mid \hmy(\bxp,\bka)+\hy(y_i,\bxp,\bka)        ,\sigma_\pi^2\right)}
       {\exp\left[\hmy(\bxp_i,\bka)+\sigma^2_\pi/2\right]\times M_y(\bka;\bxy_i,\bxp_i,\bth)}
\times p(y_i\mid \bxy_i,\bth)\nonumber\\
 =&\frac{\not\pi_i\times (1/\not\pi_i) \times\hbox{normal}(\log \pi_i\mid \hmy(\bxp,\bka)+
        \hy(y_i,\bxp,\bka),\sigma_\pi^2)}
       {\exp\left[\hmy(\bxp_i,\bka)+\sigma^2_\pi/2\right]\times M_y(\bka;\bxy_i,\bxp_i,\bth)}
\times p(y_i\mid \bxy_i,\bth),\nonumber
\end{align*}
proving the theorem.

\section{Closed-Form for $\log$ likelihood of Simple Linear Regression Model} \label{subsec:suppleLinReglikelihood}
\begin{align}
\log p_s(y_i,\pi_i\mid \bxy_i,\bxp_i,\bth,\bka)
\propto&
-\frac{1}{2} \log(\sigma_\pi^2)-\frac{1}{2\sigma_\pi^2}\left[\log \pi_i-\left(\kappa_y y_i+\bxp_i^t \bka_x   \right)\right]^2\nonumber\\
&-\frac{1}{2} \log(\sigma_y^2) -\frac{1} {2\sigma_y^2  }\left[y_i-\bxy_i^t\bbe\right]^2\nonumber\\
&-\bxp_i^t\bka_x-\sigma^2_\pi/2-  \kappa_y \bxy_i^t \bbe-\kappa_y^2\sigma_y^2/2\nonumber
\end{align}

\begin{align}
\log p_s(\by,\bpi\mid \bx,\bth,\bka)
\propto&
-\frac{n}{2} \log(\sigma_\pi^2)-
\frac{1}{2\sigma_\pi^2}\sum_i
\left[\log \pi_i-\left(\kappa_y y_i+\bxp_i^t \bka_x   \right)\right]^2\nonumber\\
&-\frac{n}{2} \log(\sigma_y^2)
-\frac{1} {2\sigma_y^2  }
\sum_i\left[y_i-\bxy_i^t\bbe\right]^2\nonumber\\
&-\sum_i \bxp_i^t\bka_x
-n \sigma^2_\pi/2
-  \kappa_y\sum_i \bxy_i^t \bbe
-n \kappa_y^2\sigma_y^2/2\nonumber
\end{align}

\subsection{Enabling Lemmas for Theorem~2} \label{AppEnabling}
We next construct two enabling results needed to prove Theorem~\ref{main} to account informative sampling under \nameref{bounded}, \nameref{independence} and \nameref{fraction}.  The first enabling result is used to bound from above the numerator in the expression for the expectation with respect to the joint distribution for population generation and the taking of the informative sample, $\left(P_{\lambda_{0}},P_{\nu}\right)$, of the fully Bayesian, sampling-weighted posterior distribution in Equation~\eqref{inform_post} on the restricted set of measures that includes those $P_{\lambda}$ that are at some \emph{minimum} distance, $\delta\xi_{N_{\nu}}$, from $P_{\lambda_{0}}$ under pseudo Hellinger metric, $d^{\pi}_{N_{\nu}}$.  The second result, Lemma~\ref{denominator}, extends Lemma $8.1$ of \citet{Ghosal00convergencerates} to bound the probability of the denominator of Equation~\eqref{inform_post} with respect to $\left(P_{\lambda_{0}},P_{\nu}\right)$, from below.  Given these $2$ Lemmas, the subsequent proof exposition for Theorem~\ref{main} is identical to \citet{savitsky2016bayesian}.

\begin{lemma}\label{numerator}
Suppose conditions~\nameref{existtests} and ~\nameref{bounded} hold.  Then for every $\xi > \xi_{N_{\nu}}$, a constant, $K>0$, and any constant, $\delta > 0$,
\begin{align}
\mathbb{E}_{P_{0},P_{\nu}}\left[\mathop{\int}_{P\in\mathcal{P}\backslash\mathcal{P}_{N_{\nu}}}\mathop{\prod}_{i=1}^{N_{\nu}}
\frac{p^{\pi}}{p_{0}^{\pi}}\left(x_{\nu i}\delta_{\nu i}\right)d\Pi\left(\lambda\right)\left(1-\phi_{n_{\nu}}\right)\right] \leq \gamma\Pi\left(\Lambda\backslash\Lambda_{N_{\nu}}\right)& \label{outside}\\
\mathbb{E}_{P_{\lambda{0}},P_{\nu}}
\left[\mathop{\int}_{\lambda\in\Lambda_{N_{\nu}}:d^{\pi}_{N_{\nu}}\left(P_{\lambda},P_{\lambda_{0}}\right)> \delta\xi}\mathop{\prod}_{i=1}^{N_{\nu}}
\frac{p^{\pi}}{p_{0}^{\pi}}\left(x_{\nu i}\delta_{\nu i}\right)d\Pi\left(\theta\right)\left(1-\phi_{n_{\nu}}\right)\right] &\leq \nonumber \\
2\gamma^{2}\exp\left(\frac{-K n_{\nu}\delta^{2}\xi^{2}}{\gamma}\right).&\label{inside}
\end{align}
\end{lemma}

This result is adjusted from that in \citet{savitsky2016bayesian} by multiplying the upper bounds on the right-hand side of both equations by constant multiplier, $\gamma \geq 1$, defined in condition~\nameref{bounded}.

\begin{proof}\label{AppNumerator}
The proof approach is the same as \citet{savitsky2016bayesian}, where we first bound the left-hand sides of Equations~\eqref{outside} and \eqref{inside} from above by an expectation of the test statistic, $\phi_{n_{\nu}}$.  When then further refine this bound on the the two subsets of measures outlined in those equations.   The refinement step is unchanged from \citet{savitsky2016bayesian}, but the first step is revised due to our unique form for our sampling-weighted, fully Bayesian estimator, outlined in Equation~\eqref{bayesrule}.  So we fully repeat the first step in the proof, below.

Fixing $\nu$, we index units that comprise the population with, $U_{\nu} = \left\{1,\ldots,N_{\nu}\right\}$.  Next, draw a single observed sample of $n_{\nu}$ units from $U_{\nu}$, indexed by subsequence, \newline$\left\{i_{\ell} \in U_{\nu}: \delta_{\nu i_{\ell}} = 1,~\ell = 1,\ldots,n_{\nu}\right\}$.  Without loss of generality, we simplify notation to follow by indexing the observed sample, sequentially, with $\ell = 1,\ldots,n_{\nu}$.

We next decompose the expectation under the joint distribution with respect to population generation, $P_{\lambda_{0}}$, and the drawing of a sample, $P_{\nu}$,

Suppose we draw $\lambda$ from some set $B \subset \Lambda$.  By Fubini,
\begin{align}
&\mathbb{E}_{P_{\lambda_{0}},P_{\nu}}\left[\mathop{\int}_{P\in B}\mathop{\prod}_{i=1}^{N_{\nu}}
\frac{p_{\lambda}^{\pi}}{p_{\lambda_{0}}^{\pi}}\left(x_{\nu i}\delta_{\nu i}\right)d\Pi\left(\lambda\right)\left(1-\phi_{n_{\nu}}\right)\right] \nonumber\\
&\leq \mathop{\int}_{\lambda\in B}\left[\mathbb{E}_{P_{0},P_{\nu}}\mathop{\prod}_{i=1}^{N_{\nu}}
\frac{p_{\lambda}^{\pi}}{p_{\lambda_{0}}^{\pi}}\left(x_{\nu i}\delta_{\nu i}\right)\left(1-\phi_{n_{\nu}}\right)\right]d\Pi\left(\lambda\right)\\
&\leq \mathop{\int}_{\lambda\in B}
\left\{\displaystyle\mathop{\sum}_{\bm{\delta}_{\nu}\in\Delta_{\nu}}\mathbb{E}_{P_{\lambda_{0}}}\left[\mathop{\prod}_{\ell=1}^{n_{\nu}}
\left[\frac{\pi_{\nu \ell}}{\pi^{\lambda}_{\nu\ell}}\frac{p_{\lambda}}{p_{\lambda_{0}}}\left(x_{\nu\ell}\right)\right]\left(1-\phi_{n_{\nu}}\right)\middle\vert \bm{\delta}_{\nu}\right]P_{P_{\nu}}\left(\bm{\delta}_{\nu}\right)\right\}d\Pi\left(\lambda\right)\label{stepdown}\\
&\leq \mathop{\int}_{\lambda\in B}
\left\{\displaystyle\mathop{\sum}_{\bm{\delta}_{\nu}\in\Delta_{\nu}}\mathbb{E}_{P_{\lambda_{0}}}\left[\mathop{\prod}_{\ell=1}^{n_{\nu}}
\left[\frac{1}{\pi^{\lambda}_{\nu\ell}}\frac{p_{\lambda}}{p_{\lambda_{0}}}\left(x_{\nu\ell}\right)\right]\left(1-\phi_{n_{\nu}}\right)\middle\vert \bm{\delta}_{\nu}\right]P_{P_{\nu}}\left(\bm{\delta}_{\nu}\right)\right\}d\Pi\left(\lambda\right)\\
&\leq \mathop{\int}_{\lambda\in B}
\mathop{\max}_{\bm{\delta}_{\nu}\in\Delta_{\nu}}\mathbb{E}_{P_{\lambda_{0}}}\left[\mathop{\prod}_{\ell=1}^{n_{\nu}}
\left[\frac{1}{\pi^{\lambda}_{\nu\ell}}\frac{p_{\lambda}}{p_{\lambda_{0}}}\left(x_{\nu\ell}\right)\right]\left(1-\phi_{n_{\nu}}\right)\middle\vert \bm{\delta}_{\nu}\right]d\Pi\left(\lambda\right)\\
&\leq \mathop{\int}_{\lambda\in B}
\mathbb{E}_{P_{\lambda_{0}}}\left[\mathop{\prod}_{\ell=1}^{n_{\nu}}
\left[\frac{1}{\pi^{\lambda}_{\nu\ell}}\frac{p_{\lambda}}{p_{\lambda_{0}}}\left(x_{\nu\ell}\right)\right]\left(1-\phi_{n_{\nu}}\right)\middle\vert \bm{\delta}^{\ast}_{\nu}\right]d\Pi\left(\lambda\right)\\
&\leq \gamma\mathop{\int}_{\lambda\in B}
\mathbb{E}_{P_{\lambda_{0}}}\left[\mathop{\prod}_{\ell=1}^{n_{\nu}}
\left[\frac{p_{\lambda}}{p_{\lambda_{0}}}\left(x_{\nu\ell}\right)\right]\left(1-\phi_{n_{\nu}}\right)\middle\vert \bm{\delta}^{\ast}_{\nu}\right]d\Pi\left(\lambda\right)\label{ratio}\\
&\leq\gamma\mathop{\int}_{\lambda\in B}\mathbb{E}_{\bm{\delta}^{\ast}_{\nu}}\left(1-\phi_{n_{\nu}}\right)d\Pi\left(\lambda\right)\nonumber,
\end{align}
where $\displaystyle\mathop{\sum}_{\bm{\delta}_{\nu}\in\Delta_{\nu}}P_{P_{\nu}}\left(\bm{\delta}_{\nu}\right) = 1$ \citep{model2003sarndal} and
\newline $\bm{\delta}^{\ast}_{\nu}\in \Delta_{\nu} = \left\{\left\{\delta^{\ast}_{\nu i}\right\}_{i = 1,\ldots,N_{\nu}},~\delta^{\ast}_{\nu i}\in\{0,1\} \right\}$ denotes that sample, drawn from the space of all possible samples, $\Delta_{\nu}$, which maximizes the probability under the population generating distribution for the event of interest.  The inequality in Equation~\eqref{ratio} results from the bound, $\frac{1}{\pi_{\nu\ell}^{\lambda}}\leq \gamma$, specified in Condition~\nameref{bounded}.

The remainder (second step) of the proof is the same as in \citet{savitsky2016bayesian}, except one multiplies their result by $\gamma$ to compute the bounds for Equations ~\eqref{outside} and \eqref{inside}.

\end{proof}

\begin{lemma}\label{denominator}
For every $\xi > 0$ and measure $\Pi$ on the set,
\begin{equation*}
B = \left\{\lambda:-\mathbb{E}_{\lambda_{0}}\log\left(\frac{p_{\lambda}}{p_{\lambda_{0}}}\right) \leq \xi^2, \mathbb{E}_{\lambda_{0}}\left(\log\frac{p_{\lambda}}{p_{\lambda_{0}}}\right)^{2} \leq \xi^{2}\right\}
\end{equation*}
under the conditions ~\nameref{sizespace}, ~\nameref{priortruth}, ~\nameref{bounded}, and ~\nameref{independence}, we have for every $C > 0 $ and $N_{\nu}$ sufficiently large,
\begin{equation}\label{denomresult}
\mbox{Pr}\left\{\mathop{\int}_{\lambda\in\Lambda}\displaystyle\mathop{\prod}_{i=1}^{N_{\nu}}\frac{p_{\lambda}^{\pi}}{p_{\lambda_{0}}^{\pi}}
\left(x_{\nu i}\delta_{\nu i}\right)d\Pi\left(\lambda\right)\leq \exp\left[-(1+C)N_{\nu}\xi^{2}\right]\right\}
\leq \frac{\gamma+C_{3}}{C^{2} N_{\nu}\xi^{2}},
\end{equation}
where the above probability is taken with the respect to $P_{\lambda_{0}}$ and the sampling generating distribution, $P_{\nu}$, jointly.
\end{lemma}

The sum of positive constants, $\gamma + C_{3}$, is greater than $1$ and will be larger for sampling designs where the (modeled) inclusion probabilities, $\{\pi_{\nu i}^{\lambda}\}$, are more variable.

\begin{proof}\label{AppDenominator}
The first part of the proof bounds the integral on the left-hand side of the event over which we take the probability   (in Equation~\eqref{denomresult}), from below, by a centered and scaled empirical process.  This first part is altered under our construction for our sampling-weighted, fully Bayesian estimator, outlined in Equation~\eqref{bayesrule}. So we specify this part, here.

By Jensen's inequality,
\begin{align*}
\log\mathop{\int}_{\lambda\in\Lambda}\mathop{\prod}_{i=1}^{N_{\nu}}\frac{p_{\lambda}^{\pi}}{p_{\lambda_{0}}^{\pi}}\left(x_{\nu i}\delta_{\nu i}\right)d\Pi\left(\lambda\right) &\geq \mathop{\sum}_{i=1}^{N_{\nu}}\displaystyle\mathop{\int}_{\lambda\in\Lambda}\log\frac{p_{\lambda}^{\pi}}{p_{\lambda_{0}}^{\pi}}\left(x_{\nu i}\delta_{\nu i}\right)d\Pi\left(\lambda\right)\\
&= N_{\nu}\cdot\mathbb{P}_{N_{\nu}}\mathop{\int}_{\lambda\in\Lambda}\log\frac{p_{\lambda}^{\pi}}{p_{\lambda_{0}}^{\pi}}d\Pi\left(\lambda\right),
\end{align*}
where we recall that the last equation denotes the empirical expectation functional taken with respect to the joint distribution over population generating and informative sampling.  By Fubini,
\begin{align}
\mathbb{P}_{N_{\nu}}\mathop{\int}_{\lambda\in\Lambda}\log\frac{p_{\lambda}^{\pi}}{p_{\lambda_{0}}^{\pi}}d\Pi\left(\lambda\right)
&= \mathop{\int}_{\lambda\in\Lambda}\left[\mathbb{P}_{N_{\nu}}\log\frac{p_{\lambda}^{\pi}}{p_{\lambda_{0}}^{\pi}}\right]d\Pi\left(\lambda\right)\\
&= \mathop{\int}_{\lambda\in\Lambda}\left[\mathbb{P}_{N_{\nu}}\delta_{\nu}
\log\left\{\frac{\pi_{\nu}}{\pi_{\nu }^{\lambda}}\frac{p_{\lambda}}{p_{\lambda_{0}}}\right\}\right]d\Pi\left(\lambda\right)\label{bayesratio}\\
&\geq \mathop{\int}_{\lambda\in\Lambda}\left[\mathbb{P}_{N_{\nu}}\frac{\delta_{\nu}}{\pi_{\nu}^{\lambda}}
\log\frac{p_{\lambda}}{p_{\lambda_{0}}}\right]d\Pi\left(\lambda\right)\label{usualratio}\\
&= \mathop{\int}_{\lambda\in\Lambda}\left[\mathbb{P}^{\pi}_{N_{\nu}}
\log\frac{p_{\lambda}}{p_{\lambda_{0}}}\right]d\Pi\left(\lambda\right)\\
&= \mathbb{P}^{\pi}_{N_{\nu}}\mathop{\int}_{\lambda\in\Lambda}\log\frac{p_{\lambda}}{p_{\lambda_{0}}}d\Pi\left(\lambda\right),
\end{align}
where we, again, apply Fubini.  Equation~\eqref{bayesratio} provides an upper bound for Equation~\eqref{usualratio} because $\pi_{\nu i} \leq 1$, $\frac{1}{\pi_{\nu i}^{\lambda}} > 0$, while $\log\frac{p_{\lambda}^{\pi}}{p_{\lambda_{0}}^{\pi}}(x_{\nu i}) < 0$.

The remainder of the proof, which uses Chebyshev to provide an upper bound on the probability of the event in Equation~\eqref{denomresult}, is specified by,
\begin{equation}\label{upperbound}
\displaystyle\leq\xi^{2}\mathop{\sup}_{\nu}\left[\frac{1}{\displaystyle\mathop{\min}_{i\in U_{\nu}}\pi_{\nu i}^{\lambda}}\right] + \xi^{2}\left(N_{\nu}-1\right)\mathop{\sup}_{\nu}\displaystyle\mathop{\max}_{i \neq j\in U_{\nu}}\left[\displaystyle\middle\vert\frac{\mathbb{E}_{\lambda_{0}}\left[\pi_{\nu ij}\right]}{\pi_{\nu i}^{\lambda}\pi_{\nu j}^{\lambda}}-1\middle\vert\right],
\end{equation}
which is nearly identical to \citet{savitsky2016bayesian} with  the raw, $\pi_{\nu i}$, replaced by the modeled, $\pi_{\nu i}^{\lambda} = \mathbb{E}_{P_{\theta}}\left(\pi_{\nu i}^{\kappa}\right)$, where $\pi_{\nu i}^{\kappa} = \mathbb{E}_{\kappa}\left(\pi_{\nu i}\vert Y_{\nu i} = y_{\nu i}\right)$ in the first term on the left of Equation~\eqref{upperbound}.  Under condition~\nameref{independence}, $\mathbb{E}_{\lambda_{0}}\left[\pi_{\nu ij}\right] \rightarrow \mathbb{E}_{\lambda_{0}}\left[\pi_{\nu i}\pi_{\nu j}\right] = \pi_{\nu i}^{\lambda_{0}}\pi_{\nu j}^{\lambda_{0}}$. Condition~\nameref{pointconverge} guarantees convergence to $1$ of the ratio in the second term of Equation~\eqref{upperbound}.
\end{proof}

\section{Data preprocessing in Section \ref{sec:application}
}\label{sub:datapreprocess}
Here we describe how we obtained our dataset of $n=6847$ participants that we used in analyses in Section
\ref{sec:application}.
We downloaded the ``Dietary Interview - Total Nutrient Intakes, First Day'' and ``Blood Pressure'' datasets.
These datasets, named DR1TOT\_H.XPT and BPX\_H.XPT, are available at
\url{https://wwwn.cdc.gov/Nchs/Nhanes/Search/DataPage.aspx?Component=Dietary&CycleBeginYear=2013}
and
\url{https://wwwn.cdc.gov/Nchs/Nhanes/2013-2014/BPX_H.htm}, respectively.

The first dataset includes the ``Dietary day one sample weight'' variable (called WTDRD1) and
its documentation states that this variable should be used as a sampling weight when analyzing it, as we did in our analyzes.
The dataset contains the nutrient intake information 9813 NHANES participants.
This dataset is derived from questionnaire data where each participant is asked questions on salt, amounts of food and beverages consumed during
the 24-hour period prior to the interview (midnight to midnight). For more details see the dataset documentation.
1152 participants have sampling weights equal to zero leaving 9813-1152=8661 (the participants with weight 0 also had missing values in all their nutrient consumption data) participants in this analysis.

\begin{figure}
\begin{center}
\includegraphics[width=55mm,angle=0]{\plotpath/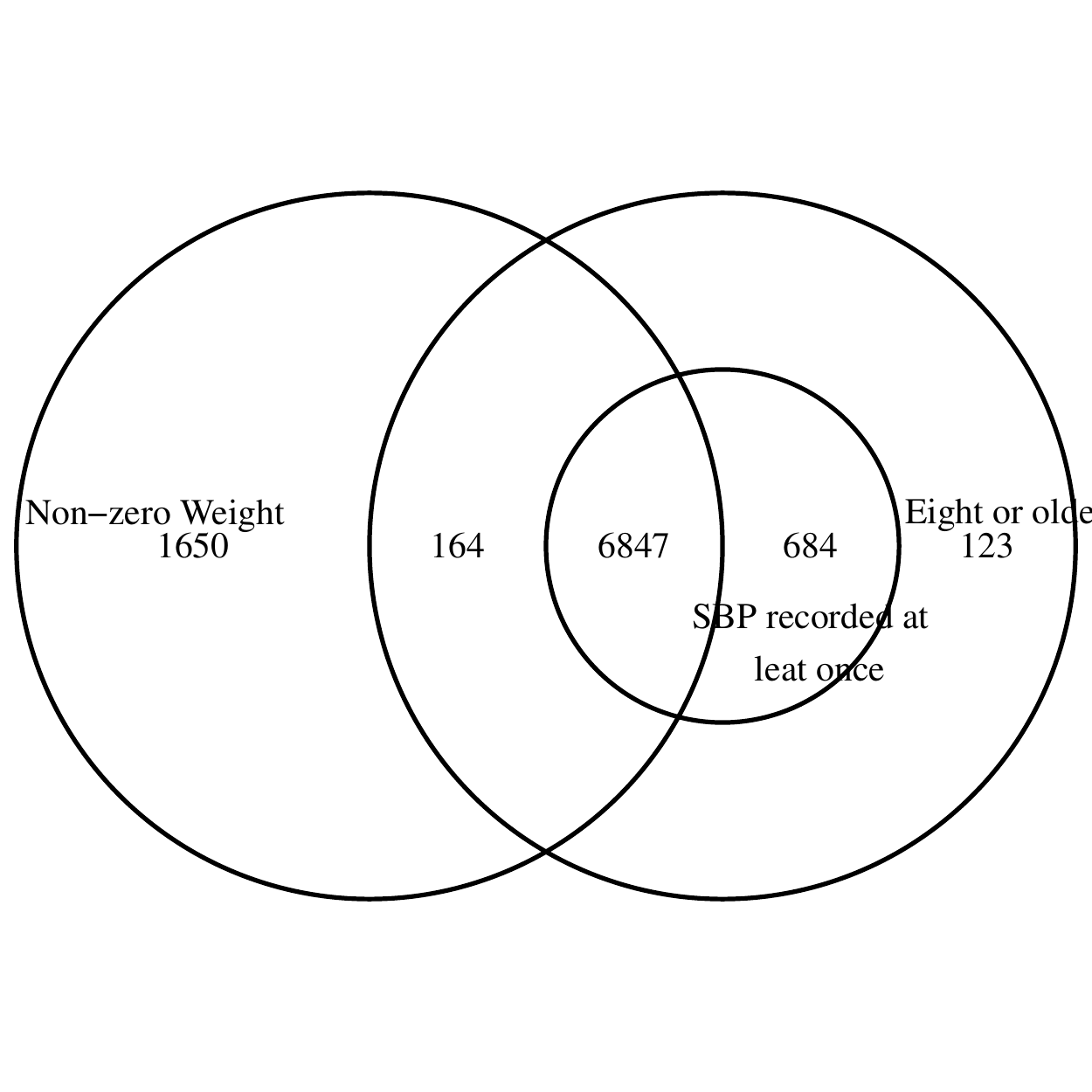} \\
\end{center}
\vskip-14mm
\caption{
\label{fig:venndiagram}
Venn diagram the three sets of NHANES participants. The sets are with non-zero weights, 8 years or older and SBP recorded at least once.
}
\end{figure}

The second dataset includes three consecutive blood pressure readings. According to its documentation,
when a blood pressure measurement is interrupted or incomplete, a fourth attempt may be made.
In our analysis SBP was the average of the not missing values of these four reads.
This dataset contains the information of the 9813 participants in the first dataset but only
 7818 are eight years or older (by design their SBP was not measured for younger participants),
 from these 807 have sampling weights equal to zero
 leaving 7818-807=7011. Out of these 7011, 6847 at least one (out of four) SBP measure recorded.
 Figure \ref{fig:venndiagram} in this Supplementary Material depicts the Venn diagram of
the sets of NHANES participants with non-zero sampling weights, 8 years or older and SBP recorded at least once.

To obtain age and gender information of the participants we downloaded the dataset
  DEMO\_H.XPT
  available at
  \url{https://wwwn.cdc.gov/nchs/nhanes/search/datapage.aspx?Component=Demographics&CycleBeginYear=2013}.
\setcounter{equation}{0}

\bibhang=1.7pc
\bibsep=2pt
\fontsize{9}{14pt plus.8pt minus .6pt}\selectfont
\renewcommand\bibname{\large \bf References}
\expandafter\ifx\csname
natexlab\endcsname\relax\def\natexlab#1{#1}\fi
\expandafter\ifx\csname url\endcsname\relax
  \def\url#1{\texttt{#1}}\fi
\expandafter\ifx\csname urlprefix\endcsname\relax\def\urlprefix{URL}\fi


\bibhang=1.7pc
\bibsep=2pt
\fontsize{9}{14pt plus.8pt minus .6pt}\selectfont
\renewcommand\bibname{\large \bf References}
\expandafter\ifx\csname
natexlab\endcsname\relax\def\natexlab#1{#1}\fi
\expandafter\ifx\csname url\endcsname\relax
  \def\url#1{\texttt{#1}}\fi
\expandafter\ifx\csname urlprefix\endcsname\relax\def\urlprefix{URL}\fi

\bibliography{mv_refs_sep2015_ss,SurveywWeights}

\begin{thebibliography}{19}
\providecommand{\natexlab}[1]{#1}
\providecommand{\url}[1]{\texttt{#1}}
\expandafter\ifx\csname urlstyle\endcsname\relax
  \providecommand{\doi}[1]{doi: #1}\else
  \providecommand{\doi}{doi: \begingroup \urlstyle{rm}\Url}\fi

\bibitem[Carpenter et~al.(2016)Carpenter, Gelman, Hoffman, Lee, Goodrich,
  Betancourt, Brubaker, Guo, Li, and Riddell]{carpenter2016stan}
Bob Carpenter, Andrew Gelman, Matt Hoffman, Daniel Lee, Ben Goodrich, Michael
  Betancourt, Michael~A Brubaker, Jiqiang Guo, Peter Li, and Allen Riddell.
\newblock Stan: {A} probabilistic programming language.
\newblock \emph{Journal of Statistical Software}, 20:\penalty0 1--37, 2016.

\bibitem[CDC-A(2016)]{CDCA}
CDC-A.
\newblock Centers for {D}isease {C}ontrol ({CDC}). national health and
  nutrition examination survey.
\newblock \url{https://www.cdc.gov/nchs/nhanes/index.htm}, 2016.
\newblock Accessed: June 7, 2016.

\bibitem[CDC-B(2016)]{CDCB}
CDC-B.
\newblock Centers for {D}isease {C}ontrol ({CDC}). modeling usual intake using
  dietary recall data, task 4: Key concepts about variance estimation methods
  in {NHANES}.
\newblock
  \url{https://www.cdc.gov/nchs/tutorials/dietary/Advanced/ModelUsualIntake/Info4.htm},
  2016.
\newblock Accessed: June 7, 2016.

\bibitem[CDC-D(2016)]{CDCD}
CDC-D.
\newblock Centers for {D}isease {C}ontrol ({CDC}). key concepts about {NHANES}
  survey design: {NHANES} sampling procedure.
\newblock
  \url{http://www.cdc.gov/nchs/tutorials/NHANES/SurveyDesign/SampleDesign/Info1.htm},
  2016.
\newblock Accessed: June 7, 2016.

\bibitem[Dong et~al.(2014)Dong, Elliott, and Raghunathan]{dong:2014}
Qi~Dong, Michael~R. Elliott, and Trivellore~E. Raghunathan.
\newblock A nonparametric method to generate synthetic populations to adjust
  for complex sampling design features.
\newblock \emph{Survey Methodology}, 40\penalty0 (1):\penalty0 29--46, 2014.

\bibitem[Ghosal et~al.(2000)Ghosal, Ghosh, and Van
  Der~Vaart]{Ghosal00convergencerates}
Subhashis Ghosal, Jayanta~K Ghosh, and Aad~W Van Der~Vaart.
\newblock Convergence rates of posterior distributions.
\newblock \emph{Annals of Statistics}, 28\penalty0 (2):\penalty0 500--531,
  2000.

\bibitem[Judkins(1990)]{judkins1990fay}
David~R Judkins.
\newblock Fay's method for variance estimation.
\newblock \emph{Journal of Official Statistics}, 6\penalty0 (3):\penalty0 223,
  1990.

\bibitem[Krewski and Rao(1981)]{krewski1981inference}
D~Krewski and JNK Rao.
\newblock Inference from stratified samples: properties of the linearization,
  jackknife and balanced repeated replication methods.
\newblock \emph{The Annals of Statistics}, pages 1010--1019, 1981.

\bibitem[Kunihama et~al.(2016)Kunihama, Herring, Halpern, and
  Dunson]{kunihama:2014}
Tsuyoshi Kunihama, AH~Herring, CT~Halpern, and DB~Dunson.
\newblock Nonparametric bayes modeling with sample survey weights.
\newblock \emph{Statistics \& Probability Letters}, 113:\penalty0 41--48, 2016.

\bibitem[Little(2004)]{Litt:mode:2004}
Roderick~J. Little.
\newblock To model or not to model? {C}ompeting modes of inference for finite
  population sampling.
\newblock \emph{Journal of the American Statistical Association}, 99\penalty0
  (466):\penalty0 546--556, 2004.

\bibitem[McCarthy(1969)]{mccarthy1969pseudo}
Philip~J McCarthy.
\newblock Pseudo-replication: Half samples.
\newblock \emph{Revue de l'Institut International de Statistique}, pages
  239--264, 1969.

\bibitem[Pfeffermann et~al.(1998)Pfeffermann, Krieger, and
  Rinott]{pfeffermann1998parametric}
Danny Pfeffermann, Abba~M Krieger, and Yosef Rinott.
\newblock Parametric distributions of complex survey data under informative
  probability sampling.
\newblock \emph{Statistica Sinica}, pages 1087--1114, 1998.

\bibitem[Pfeffermann et~al.(2006)Pfeffermann, Da~Silva~Moura, and
  Do~Nascimento~Silva]{Pfef:DaS:DoN:mult:2006}
Danny Pfeffermann, Fernando~Antonio Da~Silva~Moura, and Pedro~Luis
  Do~Nascimento~Silva.
\newblock Multi-level modelling under informative sampling.
\newblock \emph{Biometrika}, 93\penalty0 (4):\penalty0 943--959, 2006.

\bibitem[Rao and Wu(2010)]{wu:2010}
J.~N.~K. Rao and Changbao Wu.
\newblock Bayesian pseudo-empirical-likelihood intervals for complex surveys.
\newblock \emph{Journal of the Royal Statistical Society Series B}, 72\penalty0
  (4):\penalty0 533--544, 2010.

\bibitem[S\"{a}rndal et~al.(2003)S\"{a}rndal, Swensson, and
  Wretman]{model2003sarndal}
Carl-Erik S\"{a}rndal, Bengt Swensson, and Jan Wretman.
\newblock \emph{Model Assisted Survey Sampling (Springer Series in
  Statistics)}.
\newblock Springer Science \& Business Media, 2003.
\newblock ISBN 0387406204.

\bibitem[Savitsky and Toth(2016)]{savitsky2016bayesian}
Terrance~D Savitsky and Daniell Toth.
\newblock Bayesian estimation under informative sampling.
\newblock \emph{Electronic Journal of Statistics}, 10\penalty0 (1):\penalty0
  1677--1708, 2016.

\bibitem[Si et~al.(2015)Si, Pillai, Gelman, et~al.]{si2015bayesian}
Yajuan Si, Natesh~S Pillai, Andrew Gelman, et~al.
\newblock Bayesian nonparametric weighted sampling inference.
\newblock \emph{Bayesian Analysis}, 10\penalty0 (3):\penalty0 605--625, 2015.

\bibitem[Speckman and Sun(2003)]{speckman2003fully}
Paul~L Speckman and Dongchu Sun.
\newblock Fully bayesian spline smoothing and intrinsic autoregressive priors.
\newblock \emph{Biometrika}, 90\penalty0 (2):\penalty0 289--302, 2003.

\bibitem[Steffen et~al.(2012)Steffen, Kuhle, Hensrud, Erwin, and
  Murad]{steffen2012effect}
Mark Steffen, Carol Kuhle, Donald Hensrud, Patricia~J Erwin, and Mohammad~H
  Murad.
\newblock The effect of coffee consumption on blood pressure and the
  development of hypertension: a systematic review and meta-analysis.
\newblock \emph{Journal of hypertension}, 30\penalty0 (12):\penalty0
  2245--2254, 2012.

\end{thebibliography}

\bibliographystyle{plainnat}
\end{document}